\documentclass[pre,a4paper,superscriptaddress,twocolumn,amsmath,amssymb,floatfix]{revtex4}

\usepackage{graphicx}
\usepackage{amssymb}
\usepackage{float}

\usepackage{color}

\begin{document}

\title{Wigner crystal diode}

\author{Mikhail Y. Zakharov}
\affiliation{\mbox{Institute of Physics, Department of General Physics, 
Kazan Federal University, 42011 Kazan, Russia}}
\author{Denis  Demidov}
\affiliation{\mbox{Kazan Branch of Joint Supercomputer Center, 
Scientific Research Institute of System Analysis,}
\mbox{Russian Academy of Sciences, 42011 Kazan, Russia}}
\author{Dima L.~Shepelyansky}
\affiliation{\mbox{Laboratoire de Physique Th\'eorique, IRSAMC, 
Universit\'e de Toulouse, CNRS, UPS, 31062 Toulouse, France}}

\date{January 15, 2019}

\begin{abstract}
We study the transport properties of a Wigner crystal
in one- and two-dimensional asymmetric periodic
potential. We show that the Aubry transition takes place
above a certain critical amplitude of potential
with the sliding and pinned phase below and above the transition.
Due to the asymmetry the Aubry pinned phase is characterized
by the diode charge transport 
of the Wigner crystal. We argue that 
the recent experimental observations of Aubry transition
with cold ions and colloidal monolayers
can be extended to asymmetric potentials making possible
to observe Wigner crystal diode with
these physical systems 
and electrons on liquid helium.
\end{abstract}

\maketitle

\section{Introduction} 
\label{sec1}
The Wigner crystal of charged particles \cite{wigner} 
occurs when the energy of their Coulomb repulsion exceeds
the kinetic energy of their motion.
The Wigner crystal appears 
in a variety of physical systems
including electrons 
on a surface of liquid helium \cite{konobook},
electrons in two-dimensional~(2D) semiconductor samples and 
one-dimensional~(1D) nanowires 
(see e.g. \cite{matveev} and Refs. therein),
cold ions in radio-frequency traps \cite{walther,dubinrmp}
and dusty plasma in laboratory or in space \cite{fortov}.
The Wigner crystal in quasi-1D channel on 
liquid helium is also studied in experiments~\cite{kono1d}. 

It is important to understand 
how the Wigner crystal is moving
in a periodic potential in 
low-dimensional systems.
The periodic potential can be viewed
as a simplified description
of a crystal potential 
created by atoms in a solid-state system.
It also effectively appears in the frame of Little suggestion
\cite{little1,little2} on electron 
conduction in long spine conjugated polymers
with insights for possible synthesized organic
superconductors. The properties of electron
conduction in the regime of charge-density-wave (CDW)
are also related to the interacting charge
propagation in a periodic potential 
which displays a host of unusual properties \cite{thorne},
including organic superconductivity \cite{jerome,brazovski}.

The numerical and analytical analysis of
properties of 1D Wigner crystal in a periodic
potential had been started in \cite{fki2007}
with a proposal of experimental realization
of this system with cold ions in optical lattices.
It was shown that this system can be locally
described by the Frenkel-Kontorova model \cite{obraun}
where the static positions of 
interacting particles are described by the Chirikov standard map
\cite{chirikov}. This symplectic map
captures many universal features of dynamical systems
with a transition from the invariant Kolmogorov-Arnold-Moser (KAM)
curves to a global chaotic diffusion when the last KAM curve is destroyed
above the critical value of dimensional chaos parameter $K$
\cite{chirikov,lichtenberg,meiss}.
As a result this map describes behavior
of a variety of physical systems as depicted in \cite{stmapscholar}.
In this frame of dynamical systems
the irrational rotation number of the KAM curve
corresponds to the fixed incommensurate density of particles
in a periodic potential corresponding to incommensurate
crystals~\cite{pokrovsky}.

The important step in  the understanding of such incommensurate crystals
was done by Aubry \cite{aubry} showing that above the critical
value of chaos parameter $K > K_c$ the KAM curve
with an incommensurate rotation number
is replaced by an invariant Cantor set, cantori,
which has the minimal ground state energy configuration
of interacting particles in the periodic potential.
For $K<K_c$ the chain of particles
has an acoustic spectrum of low energy phonon
excitations corresponding to a sliding phase.
In contrast, for $K>K_c$ the spectrum of excitations has an optical gap
and the chain is pinned by the potential.
The properties of low energy excitations
for classical and quantum Wigner crystal
are analyzed in \cite{fki2007,snake1,snake2}  
showing the existence of exponentially many
low energy configurations in a proximity  
of the Aubry ground state. In a certain sense
for $K>K_c$ the Aubry cantori ground state, 
which is mathematically exact, is hidden inside 
exponentially large number of spin-glass-like 
configurations  which are all populated in
a physical system realization at finite temperature.
For the Wigner crystal the pinned Aubry phase
appears when the amplitude of the periodic potential exceeds 
a certain critical value measured in units of Coulomb
interaction, while at small potential amplitudes,
corresponding to the KAM curve,
the crystal can easily slide in the potential.

In addition to the very interesting fundamental 
properties of the Aubry transition from sliding to pinning,
it was established \cite{ztzs} 
that the pinned phase is characterized by the
exceptional thermodynamic characteristics 
with very large Seebeck coefficient and  figure
of merit $ZT >3$ that exceeds the largest $ZT=2.6$
value reached in material science experiments
(see e.g. review \cite{ztsci2017}).

After proposal \cite{fki2007},
the  realizations of  Wigner crystal of ions
in optical lattices attracted the interest of 
experimental groups \cite{haffner2011,vuletic2015sci}.
The signatures of the Aubry-like transition
has been experimentally detected 
with a small number of cold ions by the group of Vuletic 
with about~5 ions \cite{vuletic2016natmat}.
The chains with a larger number of cold ions
are experimentally studied in \cite{ions2017natcom}.
Two ion chains are used in  \cite{ions2017natcom}
to create an effective periodic potential 
for  ions in another chain
(zigzag transition for ions is analyzed in \cite{morigi}).
Such type of cold ion experiments can be considered
as microscopic emulators of mechanisms of nanofriction
in real materials as it is argued in \cite{tosatti1,tosatti2,tosatti3}.

Till now the cold ion traps are used to investigate 
mainly 1D or quasi-1D 
ionic Wigner crystal properties in periodic lattices.
Recently a new step to investigation of
2D Wigner crystal in  a periodic potential 
has been done with experimental observation of signatures of the
Aubry transition using a colloidal monolayer on
an optical lattice \cite{bechingerprx}.

All previous studies considered the case of periodic
potential being symmetric in space with a sine shape.
In this work we consider the case of 
asymmetric potential with two harmonics
which is relatively easy to realize 
with optical lattices (see e.g. \cite{inguscio,bloch}).
We also note that any crystal with three or more atoms in a
periodic cell in general will generate asymmetric potential.  
We show that in such a case a moderate static {\it dc}-field
$E_{dc}$ generates an asymmetric 
current corresponding to the Wigner crystal diode
with a current flowing only in one direction.
We study  this diode current for 1D case and 2D stripes
of finite width. Due to importance of diodes for electronic circuits
(see \cite{braun} and overview in \cite{horowitz})
we assume that the obtained results will clarify the
mechanisms of friction and transport on nanoscale.

\section{Model description}
\label{sec2}
For 1D case the Hamiltonian of
the system of $N$ interacting particles
with equal charges
in a periodic potential is given by
\begin{eqnarray}
\nonumber
H &=& {\sum_{i=1}^N} \big( \frac{{P_i}^2}{2} + V(x_i) \big) + U_C \; ,\\
\nonumber
 U_C &=& \sum_{i > j} \frac{ 1} {\sqrt{(x_i - x_j)^2+a^2}} \; ,\\
V(x_i) &=& K \big( \sin x_i +0.4 \sin 2x_i \big) \; .
\label{eq:ham1d}
\end{eqnarray}
Here $x_i,P_i$ are conjugated coordinate and momentum of
particle $i$, and $V(x_i)$  is an external asymmetric
periodic potential of amplitude $K$.
We use the screened Coulomb interaction $U_C$ between
particles with the screening length $a$.
Here we write the Hamiltonian in 
dimensionless units
where the lattice period $\ell=2\pi$
and the particle mass and charge are $m=e=1$.
In these atomic-like units 
the physical system parameters are measured in
units:   $r_a= \ell/2\pi$ for length,
$\epsilon_a = e^2/r_a = 2\pi e^2/\ell$ for energy,
$E_{adc} = \epsilon_a/e r_a$  
for applied static electric field,
$v_a=\sqrt{\epsilon_a/m}$ for particle velocity $v$,
$t_a =  e r_a \sqrt{m/\epsilon_a}$ for time $t$.

For 2D case the Hamiltonian has the same form with
\begin{eqnarray}
\nonumber
H &=& {\sum_{i=1}^N} \big( \frac{{P_{ix}}^2}{2} + \frac{{P_{iy}}^2}{2} 
   + V(x_i,y_i) \big) + U_C \; ,\\
\nonumber
 U_C &=& \sum_{i > j} \frac{ 1} {\sqrt{(x_i - x_j)^2+(y_i-y_j)^2+a^2}} \; ,\\
V(x_i,y_i) &=& K \big( \sin x_i +0.4 \sin 2x_i  -  \cos y_i \big) \; .
\label{eq:ham2d}
\end{eqnarray}
and 2D momentum $P_{ix}, P_{iy}$ conjugated to 
$x_i , y_i$.

Similar to \cite{ztzs} the dynamics of interacting charges
is modeled in the frame of Langevin approach (see e.g. \cite{politi})
with the equation of motion in 1D being:
\begin{equation}
\dot{P}_i = \dot{v}_i= -\partial H/\partial x_i +E_{dc} -\eta P_i+g \xi_i(t)
\; , \;\; \dot{x_i} = P_i  = v_i\; .
\label{eq:langevin}
\end{equation}
The parameter $\eta$ phenomenologically describes 
dissipative relaxation processes, and 
the amplitude of Langevin force $g$ is given 
by the fluctuation-dissipation theorem $g=\sqrt{2\eta T}$.
Here we also use particle velocities $v_i=P_i$ (since mass is unity).
As usual, the normally distributed random variables $\xi_i$ are 
defined by correlators
$\langle\langle\xi_i(t)\rangle\rangle=0$,
$\langle\langle\xi_i(t) \xi_j(t')\rangle\rangle=\delta_{i j}\delta(t-t')$.
The amplitude of the static force is given by $E_{dc}$.
For 2D case the equations of motion have the same form
with the force $E_{dc}$ acting in $x$-direction.

The length of the system in 1D case is taken to
be $2\pi L$ in $x$-direction
with $L$ being the integer number of periods
with periodic boundary conditions. 
In 2D we studied the case of stripes with
the width of $2\pi N_s$ considering usually up to 
$N_s=5$ period cells in $y$-direction with periodic boundary conditions.

The numerical simulations are based on the combination of
Boost.odeint~\cite{odeint} and VexCL~\cite{demidov2013, vexcl} libraries and
employed the approach described in \cite{ahnert2014} in order to accelerate the
solution with NVIDIA CUDA technology.  Problem ~\eqref{eq:ham1d}
are solved using fourth order Runge-Kutta method and~\eqref{eq:ham2d} 
by Verlet method, where
each particle is handled by a single GPU thread.  Since Coulomb interactions
in $U_C$ are decreasing with distance between particles, 
the interactions for the 2D case were cut off at the
radius $R_C=6\ell = 12\pi$, that allowed to reduce the computational complexity of the
algorithm from $O(N^2)$ to $O(N \log N)$.  In 1D in some cases we only
considered interactions between immediate left and right neighbors, since, as
shown in \cite{fki2007,ztzs}, the contribution of other particles does not play
a significant role.  In all simulations, in order to avoid close encounters
between particles leading to numerical instability, the screening length
$a=0.7$ is used.  At this value of $a$ the interaction energy is still
significantly larger than the typical kinetic energies of particles ($T \ll 1/a$) 
and the screening does not significantly affect the interactions of
particles.  We usually employed the relaxation rate $\eta=0.1$ being relatively
small comparing to typical oscillation time scales in the system, but other
values of $\eta$ were also considered.  The source code for the 1D and 2D
experiments is available at https://gitlab.com/ddemidov/thermoelectric1d and
https://gitlab.com/ddemidov/thermoelectric2d correspondingly.  The numerical
simulations were run at OLYMPE CALMIP cluster \cite{olympe} using NVIDIA Tesla
V100 GPUs and partially at Kazan Federal University using NVIDIA Tesla C2070
GPUs. 

\section{Static configurations and Aubry transition}
\label{sec3}

\begin{figure}[t]
\begin{center}
\includegraphics[width=0.48\textwidth]{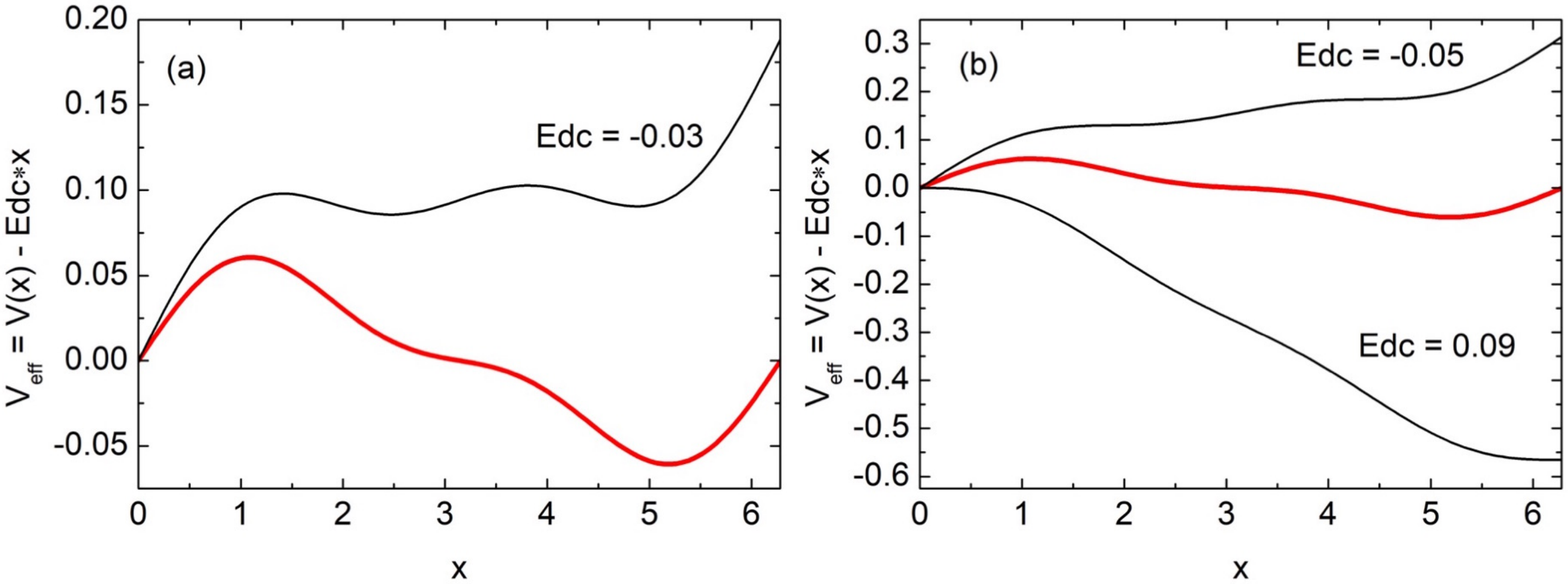}
\end{center}
\vskip -0.3cm
\caption{\label{fig1}
The effective static potential $V_{eff}(x)=V(x) - E_{dc} x$ for one charge
is shown by black and red/gray curves
in presence of a static force $E_{dc}$ at $K=0.05$,
red/gray curve corresponds to $E_{dc}=0$;
black curves show nonzero values of $E_{dc}$
in panels (a), (b).
}
\end{figure}

We start from the analysis of properties of static 
configurations of particles in 1D case. The effective
potential $V_{eff}=V(x)-E_{dc} x$,
acting on a particle in presence of a static force $E_{dc}$, 
is shown in Fig.~\ref{fig1}. 
At $E_{dc}=0$ there is one potential minimum.
In presence of the static force $E_{dc}$ there are no potential 
minima for the force $E_{dc} > -K$ pushing to the left
and $E_{dc} < 1.8 K$ pushing to the right.
Thus the asymmetry of potential leads to different
sliding borders for left and right force acting on one particle. 
This asymmetry is at the origin of diode transport
of Wigner crystal.

At $E_{dc}=0$
the static configurations of 
Wigner crystal energy local minima
are defined by the conditions $\partial H/ \partial x_i =0$
\cite{fki2007,obraun,aubry}.
As discussed in \cite{fki2007},
in the approximation of nearest neighbor
interacting charges, these conditions
lead to the dynamical symplectic Wigner map 
for equilibrium positions $x_i$ of charges in the Wigner crystal
(with $P_i=v_i=0$):
\begin{eqnarray}
p_{i+1} = p_i + K g(x_i) \; , \; \; x_{i+1} = x_i+1/\sqrt{p_{i+1}} \; ,
\label{eq:map}
\end{eqnarray}
where the effective momentum conjugated to $x_i$ is
$p_i = 1/(x_{i}-x_{i-1})^2$ and the kick function
$K g(x)= -dV(x)/dx = -K (\cos x +0.8 \cos 2x)$.

To check the validity of the map description we
find the ground state configuration using numerical methods
of energy minimization 
described in \cite{fki2007,aubry}.  
Here the Coulomb interaction between all electrons 
is used in the numerical simulations.
We use the hard wall boundary conditions
at the ends of the chain (for ions they  can
be created by specific laser frequency detuning from
resonant transition between ion energy levels).
This leads to the density $\nu$ of charges along the chain
being  inhomogeneous since a charge
in a boundary vicinity has more pressure from other
charges in the chain (a similar inhomogeneous 
local density $\nu(x_i) =2\pi/\mid x_{i+1} - x_i \mid$
appears for ions inside a global oscillator 
potential of a trap as discussed in \cite{fki2007}).
Thus, as in \cite{fki2007}, we select the central
part of the chain with approximately $1/3$ of all charges
where the density is approximately constant 
being close to the golden mean value
$\nu = \nu_g - 1 =(\sqrt{5}-1)/2 \approx 0.618$ 
(or $\nu = \nu_g= (\sqrt{5}+1)/2  \approx 1.618$)
which is assumed to be a most robust
KAM curve for the Chirikov standard map~\cite{lichtenberg,meiss,aubry}.
This choice corresponds to an
incommensurate phase with the golden KAM curve 
usually studied for the Aubry transition \cite{lichtenberg,meiss,aubry}.
 
The numerically obtained charge positions and momentum $x_i,p_i$
are shown in Fig.~\ref{fig2} for $\nu \approx 0.618$
(see Fig.~\ref{figA1} in  Appendix for $\nu \approx 1.618$).
From the numerical values $x_i$ we determine 
the kick function $g(x)$ which is close to the 
theoretically expected relation $Kg(x)=-dV/dx$
shown by the dashed curve.
For small potential amplitudes $K<0.0015$
the chain is in the sliding
phase with a continuous KAM curve in the 
Poincar\'e section plane $(x,p)$.
For  $K>0.0015$ the points $(x_i,p_i)$
start to be embedded inside the chaotic component of the phase plane
corresponding to the Aubry pinned phase.
In this phase the points  $(x_i,p_i)$
form a fractal Cantor set in the phase plane
and the chain  is pinned by the potential. 
According to the data of Fig.~\ref{fig1}
the Aubry transition from sliding to pinned phase
takes place at the critical value $K_c \approx 0.0015$. 
The qualitative change of chain properties
with the transition from sliding to pinned
phase can be also seen with the help of hull function
$h(x)$ which gives the charge
positions in a periodic potential vs. unperturbed positions
at $K=0$ both taken $\mod 2\pi$.
For $K<K_c$ we have a continuous function $h(x) \approx x$ while for
$K>K_c$ the hull function has a form of devil's staircase 
with charge positions clustering near certain values.
For $\nu \approx 1.618$ we find $K_c \approx 0.015$
(see Fig.~\ref{figA1} in Appendix).

For the potential $V(x) = - K \cos x$ the Wigner map~(\ref{eq:map})
can be locally described by the Chirikov standard map
as it is explained in \cite{fki2007}. This gives the 
Aubry transition at the potential amplitude 
\begin{eqnarray}
K_{c\nu} \approx 0.034 (\nu/\nu_g)^3 \; , \;\; \nu_g=1.618... \; .
\label{eq:kc}
\end{eqnarray} 
At $\nu=1.618...$ the numerical results
obtained in \cite{fki2007,ztzs} give $K_c \approx 0.0462$
that is slightly  above the theoretic value.
We attribute this modest difference to 
an inhomogeneous density of resonances in (\ref{eq:map})
(the Chirikov standard map approximation assumes it
to be constant~\cite{chirikov}) and to nearest neighbor interactions
between charges present in the Wigner crystal.
The recent results confirm the cubic decrease
of $K_{c\nu}$ with charge density $\nu$ \cite{ztions}.

\begin{figure*}[t]
\begin{center}
\includegraphics[width=0.9\textwidth]{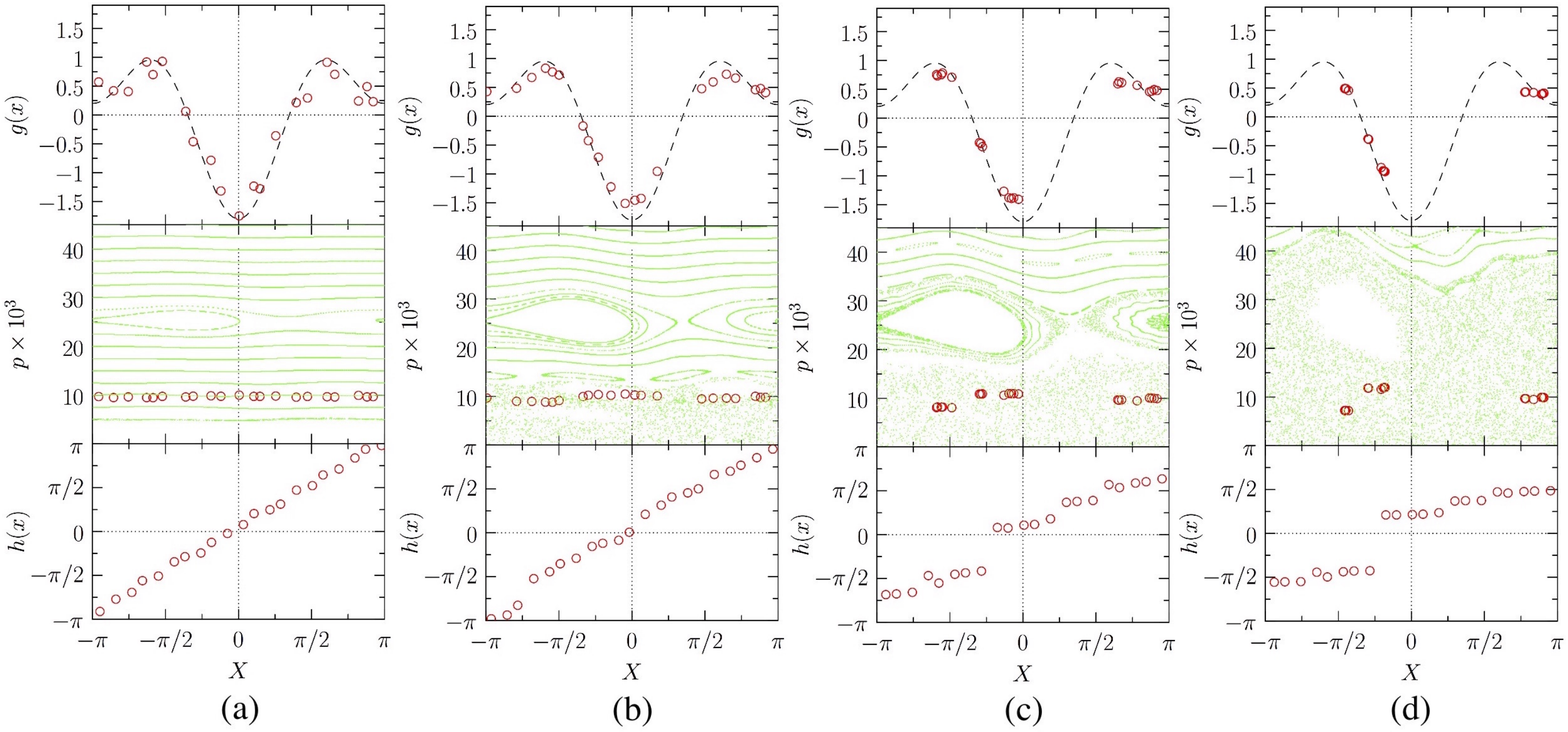}
\end{center}
\caption{\label{fig2}
Functions related to the dynamical map (\ref{eq:map})
obtained from the ground state equilibrium positions $x_i$ of $N=61$ 
charged particles for $L=90$ potential periods 
(with hard wall boundary conditions) 
for density $\nu \approx 0.618$ in the central chain part 
at $K=0.0002$ (a), $K=0.001$ (b), $K=0.002$ (c),
$K=0.005$ (d) (charges are marked by open circles). 
In each panel the top subpanel shows the kick function
$g(x)$ (dashed curve is the theoretical curve, circles are actual
charge positions); middle subpanel shows the Poincar\'e section
of map (\ref{eq:map}) (green/gray points) and actual charge positions
$(x_i,p_i)$ (open circles);
bottom subpanel shows the hull function $h(x)$ (see text).
The charge positions
are shown as $x=x_i (\mod 2\pi)$ for the central 
1/3 part of the chain. 
}
\end{figure*}

\newpage
In our case with two harmonics of potential
the density of resonances is increased 
which, according to the Chirikov criterion
of overlapped resonances, should
decrease $K_c$ value (see \cite{chirikov,lichtenberg}).
Indeed, for $\nu =1.618$ we have $K_c \approx 0.015$ being approximately
$3$ times smaller compared to the case of one harmonic
potential with $K_c \approx 0.0462$.
For $\nu \approx 0.618$ we have $K_{c\nu} \approx 0.0015$
while the cubic extrapolation like in (\ref{eq:kc})
gives $K_{c\nu} \approx 0.00084$ being by 40\% lower than the numerical 
values from Fig.~\ref{fig2}. We consider this 
as satisfactory taking into account
the approximate $K_c$ values extracted from Fig.~\ref{fig2} 
and Fig.~\ref{figA1}. Thus for the potential (\ref{eq:ham1d})
we have on average the density dependence of the critical
potential amplitude of the Aubry transition:
$K_{c\nu} \approx 0.01 (\nu/\nu_g)^3$.

We note that the significant decrease of $K_c$
for two harmonic potential and especially with density $\nu$
can be rather important for experiments with optical 
lattices since a smaller potential amplitude
is more accessible with low power lasers.

\section{Diode transport in 1D}
\label{sec4}

\begin{figure}[t]
\begin{center}
\includegraphics[width=0.49\textwidth]{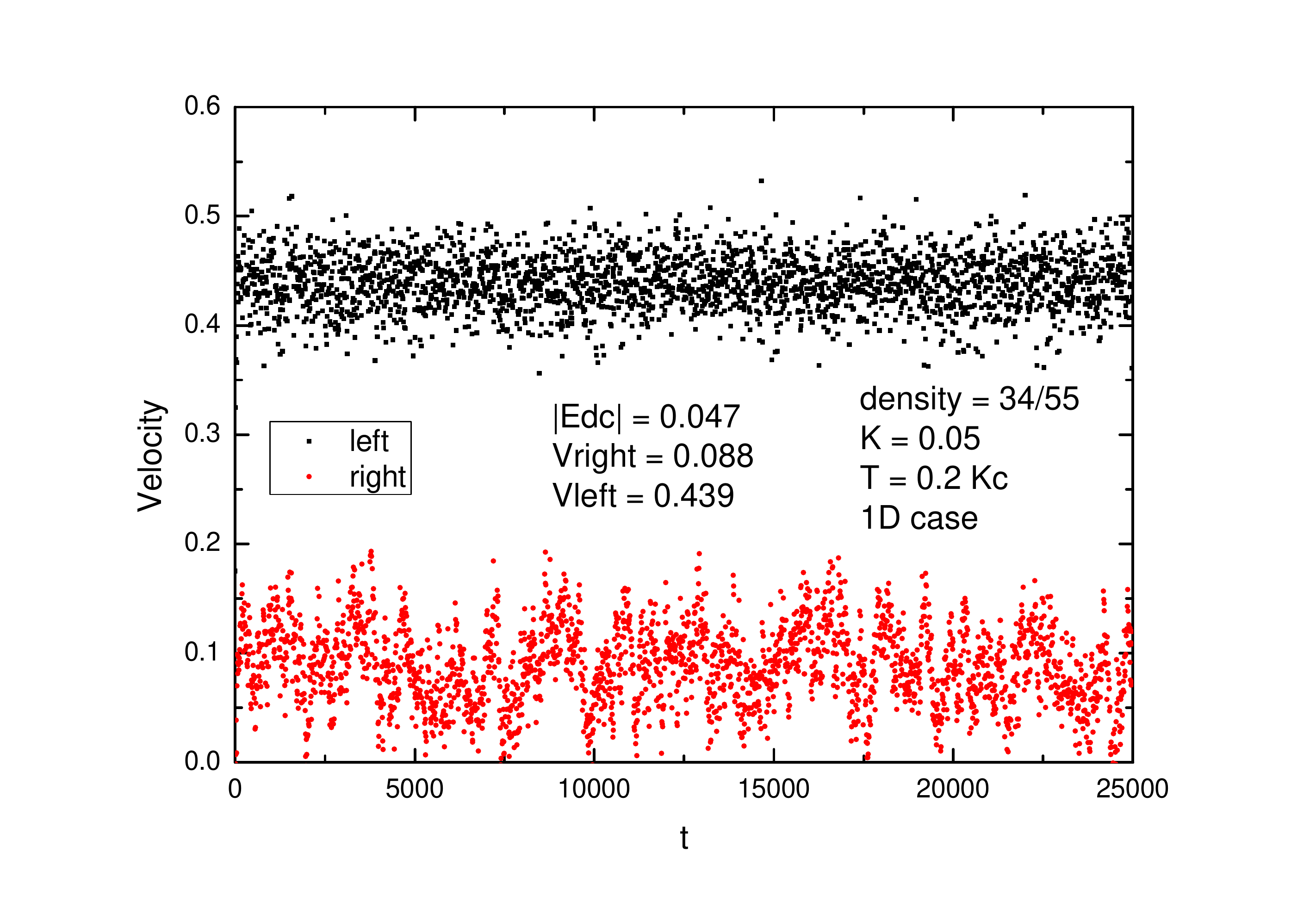}
\end{center}
\vskip -0.3cm
\caption{\label{fig3}
Dependence of average chain velocity $v(t)$ at time $t$ in 1D
for $N=34$ particles and $L=55$ periods of potential
(with periodic boundary conditions; $\nu=34/55$).
Here $E_{dc}=-0.047$ (black points) with time average 
Wigner crystal velocity
$v_W=v_{left}=-0.439$ and $E_{dc}=0.047$ (red/gray points)
with time average $v_W=v_{right}=0.088$;
the system parameters are
$K=0.05, T=0.2K_c , K_c=0.0462$; $\eta=0.1$
(time averaging is done over times above 1/2
of the whole time interval). 
}
\end{figure}

The charge transport is computed with $N$ charges on $L$ potential periods
with periodic boundary conditions (here we consider Coulomb interactions
only between nearest neighbor charges as discussed in Section ~\ref{sec2}). 
We compute the velocity $v(t)$ of a chain at time $t$
as an average velocity of all $N$ charges at that time moment.
A typical dependence of $v(t)$ on time is shown in Fig.~\ref{fig3}.
The system parameters correspond to the Aubry pinned phase.
We see that approximately for $t > 100$ the motion of a chain
is in a stationary regime with its steady-state propagation
along the potential under the applied static force $E_{dc}$.
The velocity of Wigner crystal propagation $v_W$ to the left
is close to the velocity of a free particle 
in presence of force and dissipation $v_0=E_{dc}/\eta =0.47$
for the case of Fig.~\ref{fig3}. The time averaged value
$v_{left} =0.439$ is a bit smaller than $v_0$ showing that 
the potential slightly decreases the propagation velocity.
In contrast, the chain propagation to the right
has a significantly smaller velocity
$v_{right} =0.089 \ll v_0$ for the same amplitude of static force.
Also instantaneous values of velocity $v(t)$ have significant 
fluctuations with even almost zero velocity at some moments of time.
This data clearly demonstrates the emergence of
diode transport in the asymmetric potential in presence of interactions.
From the physical view point the velocity
to the right is smaller than to the left
since the potential has a steep slope in this direction while
moving to the left a particle follows a gentle slope.
Thus in winter it is easier to pull sleigh
along a gentle slope of a hill
than along a steep slope even if the hill height is the same
from both sides.

\begin{figure}[t]
\begin{center}
\includegraphics[width=0.47\textwidth]{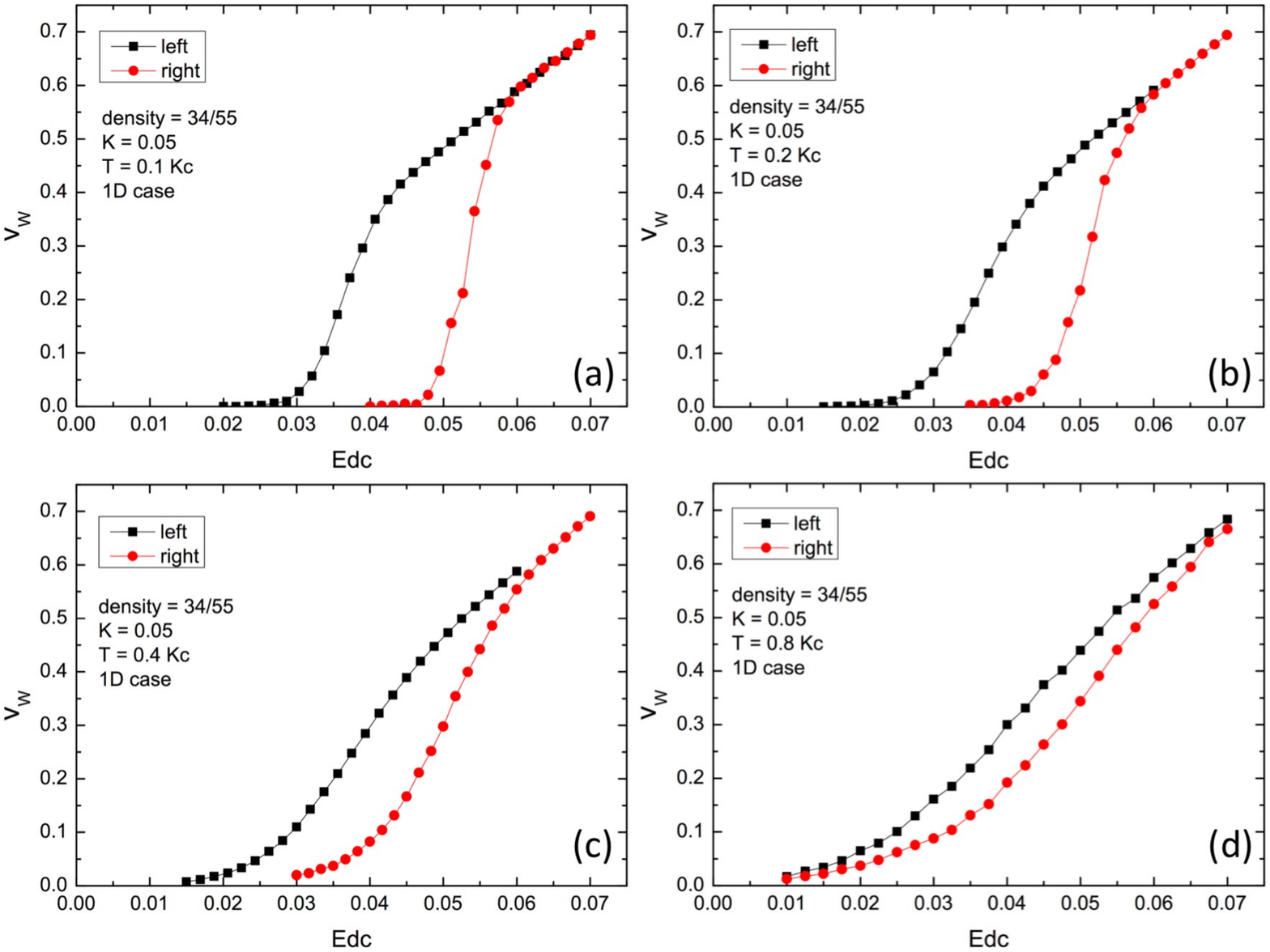}
\end{center}
\vskip -0.3cm
\caption{\label{fig4}
Dependence of Wigner crystal velocity $v_W$, averaged over times $t \leq 10^4$, 
on static field amplitude $E_{dc}$ for propagation to the left
(black points) and to the right (red/gray points).
Here $\nu=N/L=34/55$, $K=0.05$, $\eta=0.1$
and temperature $T/K_c=0.1; 0.2; 0.4; 0.8$ 
for panels (a); (b); (c); (d) respectively.
Here $K_c=0.0462$ is the critical amplitude of Aubry transition
for $\nu=1.618...$ in a potential with one harmonic \cite{fki2007,ztzs}. 
}
\end{figure}

\begin{figure}[t]
\begin{center}
\includegraphics[width=0.49\textwidth]{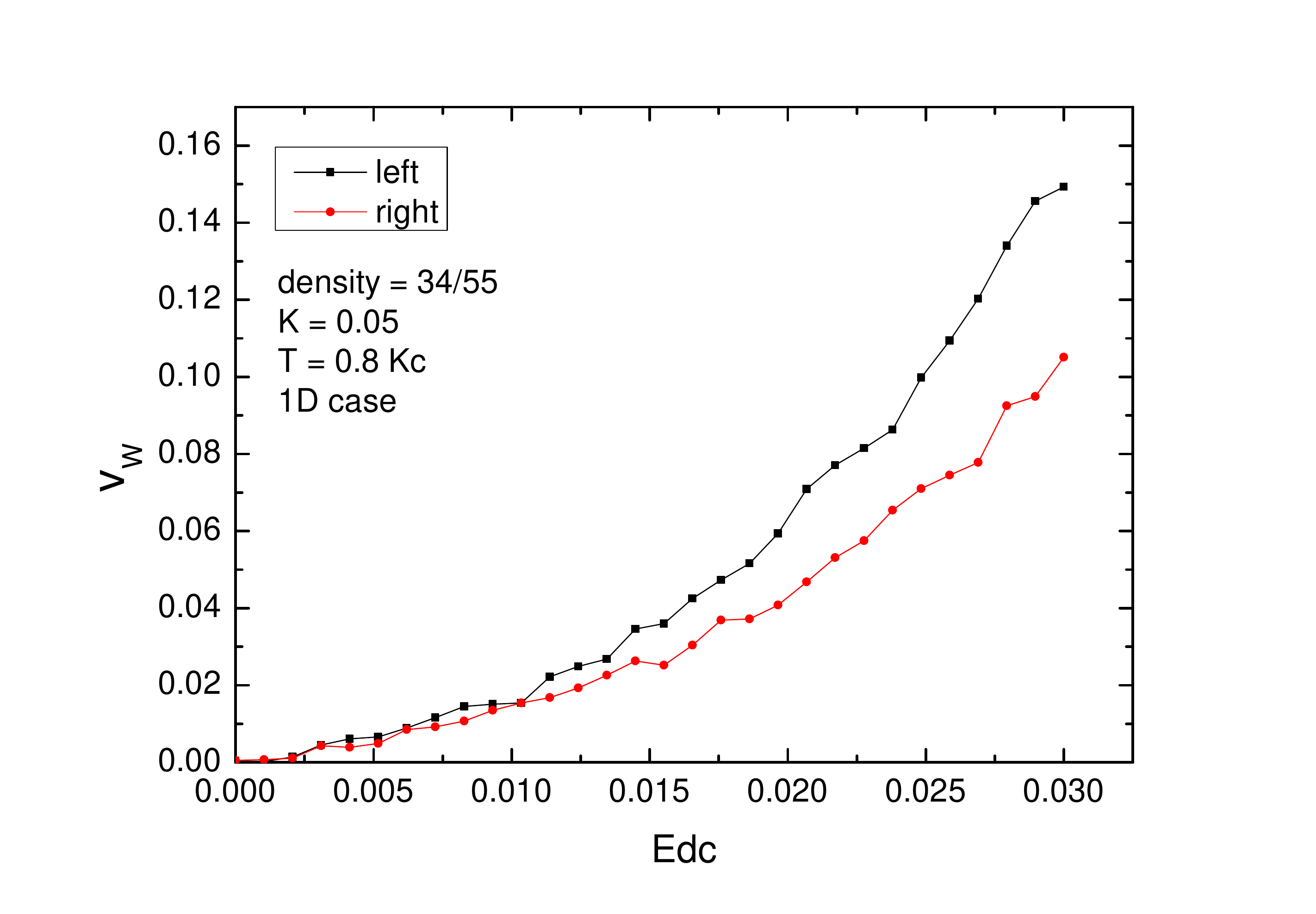}
\end{center}
\vskip -0.3cm
\caption{\label{fig5}
Zoom of Fig.~\ref{fig4}(d) at small $E_{dc}$.
}
\end{figure}

The dependence of Wigner crystal velocity $v_W$ on
$E_{dc}$ is shown in Fig.~\ref{fig4}
for the Aubry pinned phase at $K=0.05$
and $\nu =N/L \approx 0.618$
for different values of temperature~$T$.
At small $T = 0.1 K_c =0.00462 \ll K=0.05$
we have a strongly asymmetric diode transport
appearing at finite $E_{dc}$  fields. With the increase
of temperature the diode transport becomes less and less pronounced.
Indeed, when the temperature is comparable with the
potential height, e.g. $T=0.8Kc \approx 0.037 \sim K =0.05$,
the statistical Boltzmann fluctuations 
over potential barrier smooth the asymmetry of transport.
In this regime at small $E_{dc}$ fields the velocities
to the left and to the right directions become the same
as it is shown in Fig.~\ref{fig5}. 

Indeed, in the linear regime 
limit at $E_{dc} \rightarrow 0$
the principle of detailed balance (see e.g. \cite{landau10}) guaranties
that the flow is the same in both directions.
We note that this point had been discussed in detail by Feynman for 
a case of asymmetric potential \cite{feynman} which became known as  ratchet.
Indeed,  as we show below (see Section~\ref{sec7})
at small temperatures the dependence $v_W(T)$
is well described by the Arrhenius thermal activation equation.
At present, the term {\it ratchet}, discussed by Feynman, 
is more used for a description of
directed transport appearing in an asymmetric periodic potential
under a time-periodic force driving (see e.g. \cite{prost,hanggi,entin}).
Due to these reasons we use the term diode 
which is more adequate for the case of static force
without any time-periodic driving.
It directly corresponds to the asymmetry of charge flow 
obtained in our model.

\begin{figure}[t]
\begin{center}
\includegraphics[width=0.49\textwidth]{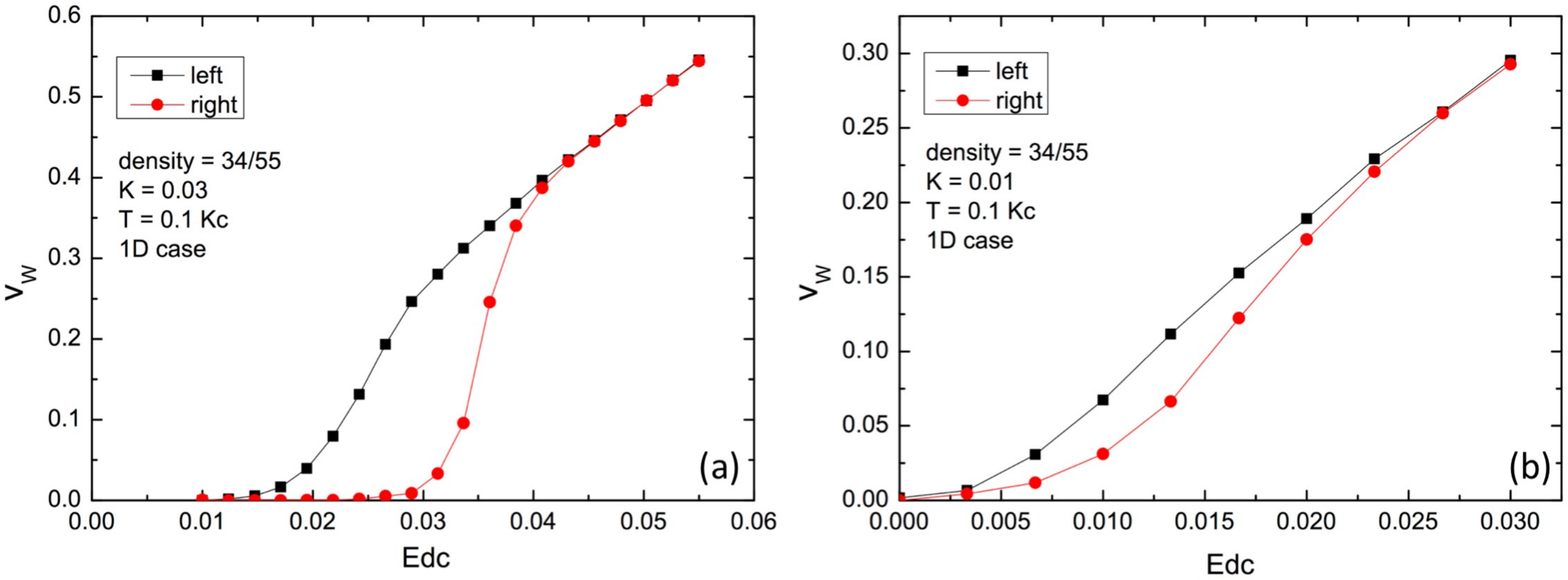}
\end{center}
\vskip -0.3cm
\caption{\label{fig6}
Same as in Fig.~\ref{fig4}(a) but
at $K=0.03$ (a) and $K=0.01$ (b).
}
\end{figure}

At fixed temperature (e.g at $T=0.1K_c$ as in Fig.~\ref{fig4}(a))
the asymmetry of diode transport decreases with a decrease of
the potential amplitude $K$ as it is shown in Fig.~\ref{fig6}.
Indeed, with a decrease of $K$ the potential height becomes
comparable with temperature and, as above,
the principle of detailed balance
leads to the same flows in both directions.

\begin{figure}[t]
\begin{center}
\includegraphics[width=0.48\textwidth]{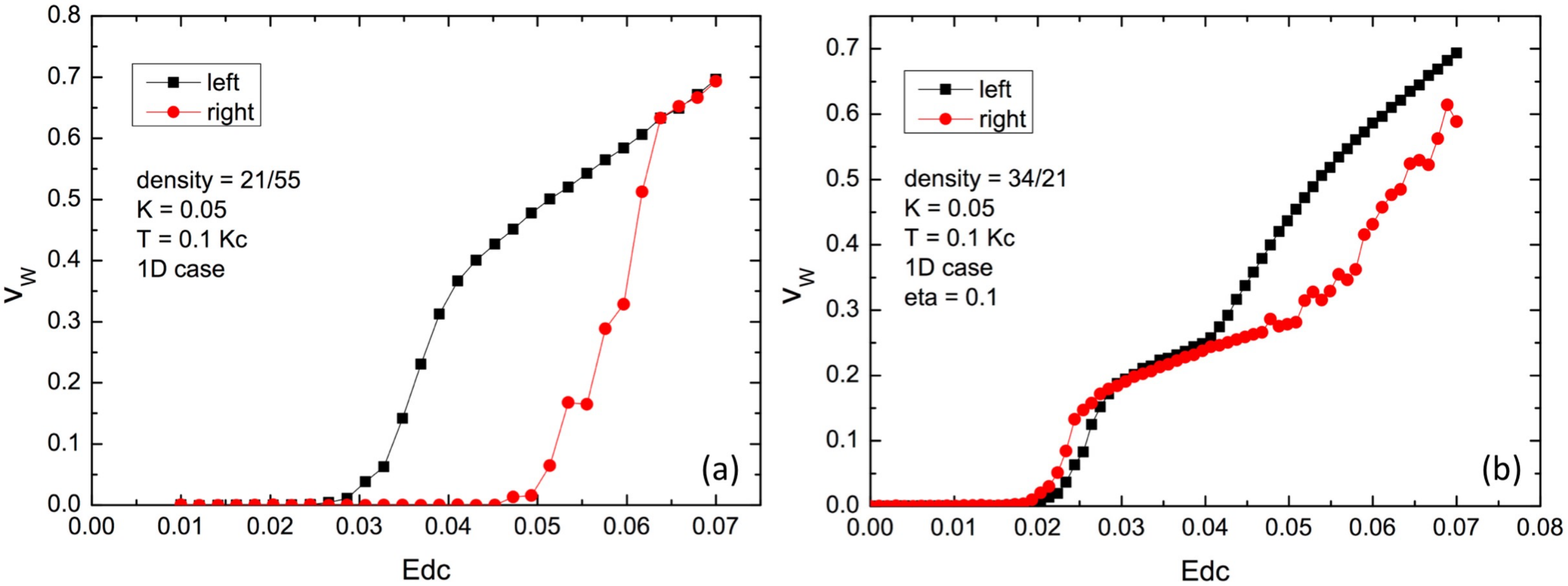}
\end{center}
\vskip -0.3cm
\caption{\label{fig7}
Same as in Fig.~\ref{fig4}(a)
but for different electron densities $N/L=21/55$(a)
and $N/L=34/21$(b). 
}
\end{figure}

The influence of Coulomb interactions
of the diode transport is obtained 
from the comparison of Fig.~\ref{fig7}(a),
Fig.~\ref{fig4}(a), Fig~\ref{fig7}(b)
where the electron density takes values 
$\nu=N/L = 21/55, 34/55, 34/21$ respectively,
at all other parameters kept fixed.
All these 3 cases are located in 
the pinned phase. We see that at 
$N/L=21/55, 34/55$ the diode velocity dependence
$v_W(E_{dc})$ on $E_{dc}$ remains practically the same.
We explain this by the fact that 
at small densities the interactions
between charges become weak compared
to the periodic potential 
and thus we approach to the limit of 
transport of noninteracting  particles
which still demonstrates the diode flow
due to potential asymmetry.
The opposite tendency appears with the increase of density
to $N/L=34/21$. In this case the interactions are 
strong, even if we are still in the pinned phase,
and the asymmetry of potential is less pronounced
so that the transport flows are close to be symmetric
even if one still needs to apply a finite force $E_{dc} \approx 0.02$
to obtain  moderate velocities of Wigner crystal motion
along the periodic potential.   This force can be interpreted as 
the static friction force $F_s \approx E_{dc} \approx 0.02$. 
We note that with the increase
of interactions (or density)
we have a reduction of $F_s$ values.
Indeed, at lower density $N/L=34/55$ we have 
larger values of static friction force $F_s \approx 0.03$ 
(for left direction)
and $F_s \approx 0.05$ (for right direction).
Indeed, with increase of interactions
we approach to the sliding phase 
(the border of Aubry transition is growing this density 
(\ref{eq:kc}) where $F_s =0$.

We note that for $E_{dc}   < F_s$ the Wigner crystal velocity $v_W$
happens due to the Arrhenius thermal activation
(see Section~\ref{sec7}) so that it drops
exponentially with a decrease of temperature.
In contrast, for $E_{dc} > F_s$ we have rather weak change
of $v_W$ with $T$ (see e.g. Fig.~\ref{fig4}).

The above results are presented for the dissipation
rate $\eta=0.1$. We tested also smaller values of $\eta$
for which we obtained qualitatively similar results.
Thus for parameters of Fig.~\ref{fig7}(b) we 
have approximately similar shape of $v_W(E_{dc})$ dependence
for $\eta=0.1$ and $\eta=0.02$ (see Fig.~\ref{figA2})
with a more sharp shape in the latter case
with a smaller value $F_s \approx 0.013$.
We argue that at smaller dissipation 
statistical fluctuations 
at a given temperature have more possibilities 
to overcome potential barriers that leads to a 
moderate decrease of $F_s$.

\section{Diode transport in 2D}
\label{sec5}

The diode properties of Wigner crystal in 2D
are studied with the Hamiltonian (\ref{eq:ham2d})
and related Langevin equations (\ref{eq:langevin}).
We use periodic boundary conditions in $x$ and $y$ directions
with the static field $E_{dc}$ always acting in $x$-direction.
The majority of the results were obtained with $5$ cells in $y$
giving us $N_s=5$ stripes along $x$ with one periodic cell in each stripe
(we obtained very similar results with only $N_s=1$ stripe in $y$).
We keep the same density $\nu=N/L$ in each stripe as 
in the above 1D case. Thus the total number of charges is $N_{tot}=N_s N$
with $L$ and $N_s$ potential periods in $x$ and $y$.
Similar to 1D case (see Fig.~\ref{fig2})  
we compute the local average charge velocity $v(t)$
in $x$-direction at instant time moment $t$  by
averaging over all $N_{tot}$ charges (an example of dependence 
$v(t)$ is shown in Fig.~\ref{figA3}). 
In 2D simulations we usually used time scales up to $t =5 \cdotp 10^4$ 
when the crystal propagation is well stabilized.

\begin{figure}[t]
\begin{center}
\includegraphics[width=0.48\textwidth]{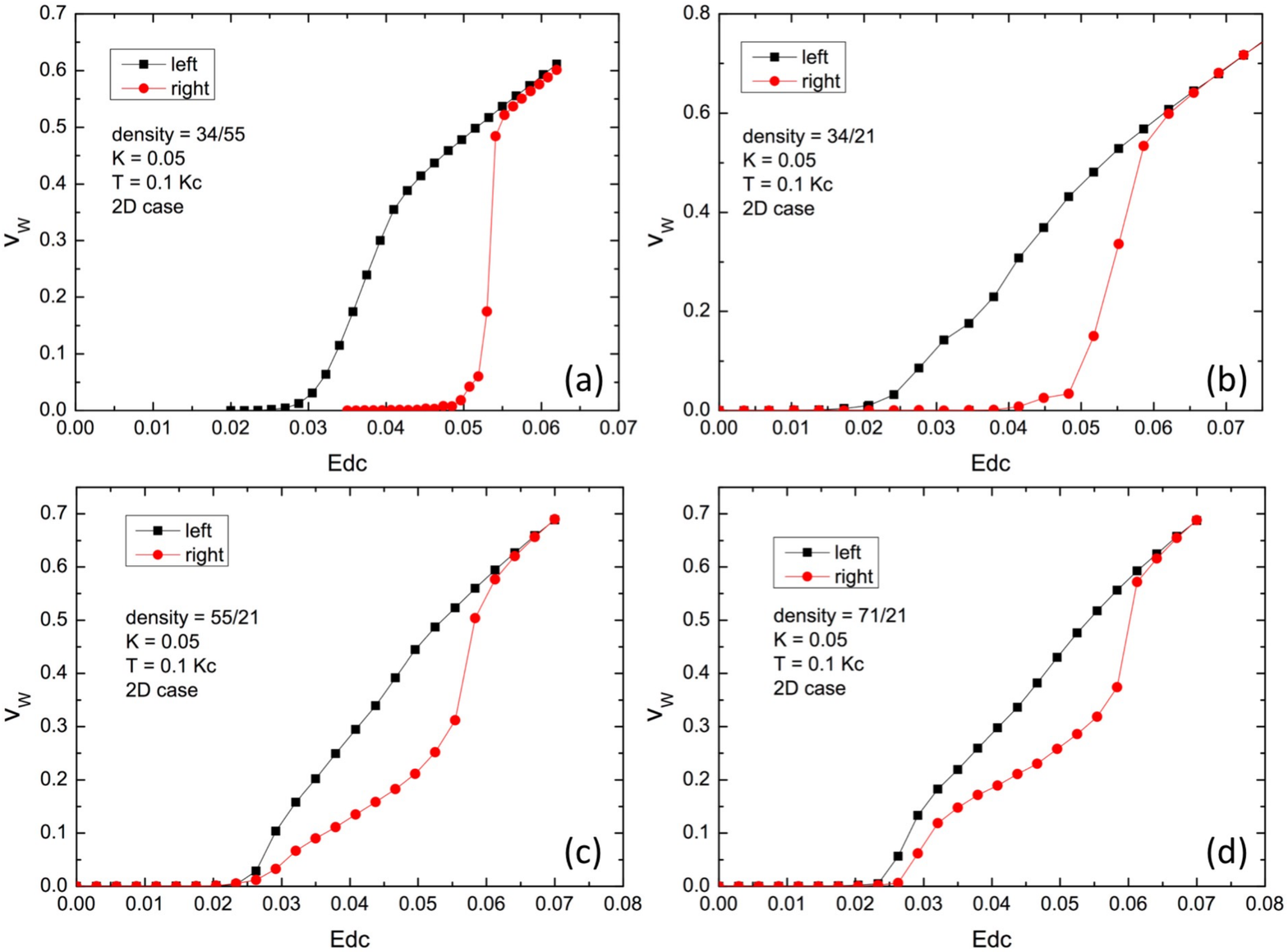}
\end{center}
\vskip -0.3cm
\caption{\label{fig8}
Dependence of Wigner crystal velocity $v_W$ on
$E_{dc}$ in 2D case (5 lines in transverse $y$ direction)
at densities $\nu=N/L = 34/55$(a), $34/21$(b),
$55/21$(c), $71/21$(d). Other system
parameters are as in Fig.~\ref{fig4}(a).
}
\end{figure}

The dependence of Wigner crystal velocity $v_W$
of $E_{dc}$ in 2D is shown in Fig.~\ref{fig8}
at difference densities $\nu = N/L = 34/55, 34/21, 55/21, 71/21$
with all other parameters kept fixed
being also the same as in the corresponding 1D cases.
At low density $N/L=34/55$ the dependence 
$v_W(E_{dc})$ remains practically the same in 1D and 2D.
We attribute this to relatively weak interactions
between charges so that the diode transport is rather
close to the noninteracting case. The situation
is drastically different for $N/L=34/21$:
in 1D the diode transport asymmetry is quite weak (Fig.~\ref{fig7}(b))
while in 2D we have a strong asymmetry (Fig.~\ref{fig8}(b))
with different values of static friction force
$F_s \approx 0.025$ (left) and $F_s \approx 0.05$ (right). 
At larger densities $N/L=55/21, 71/21$ the asymmetry
is reduced due to increase of interactions
with $F_s \approx 0.025$ for both directions.
Thus in 2D case we have a tendency similar to 1D case
with a reduction of diode transport asymmetry 
with the density increase. 

\begin{figure}[t]
\begin{center}
\includegraphics[width=0.49\textwidth]{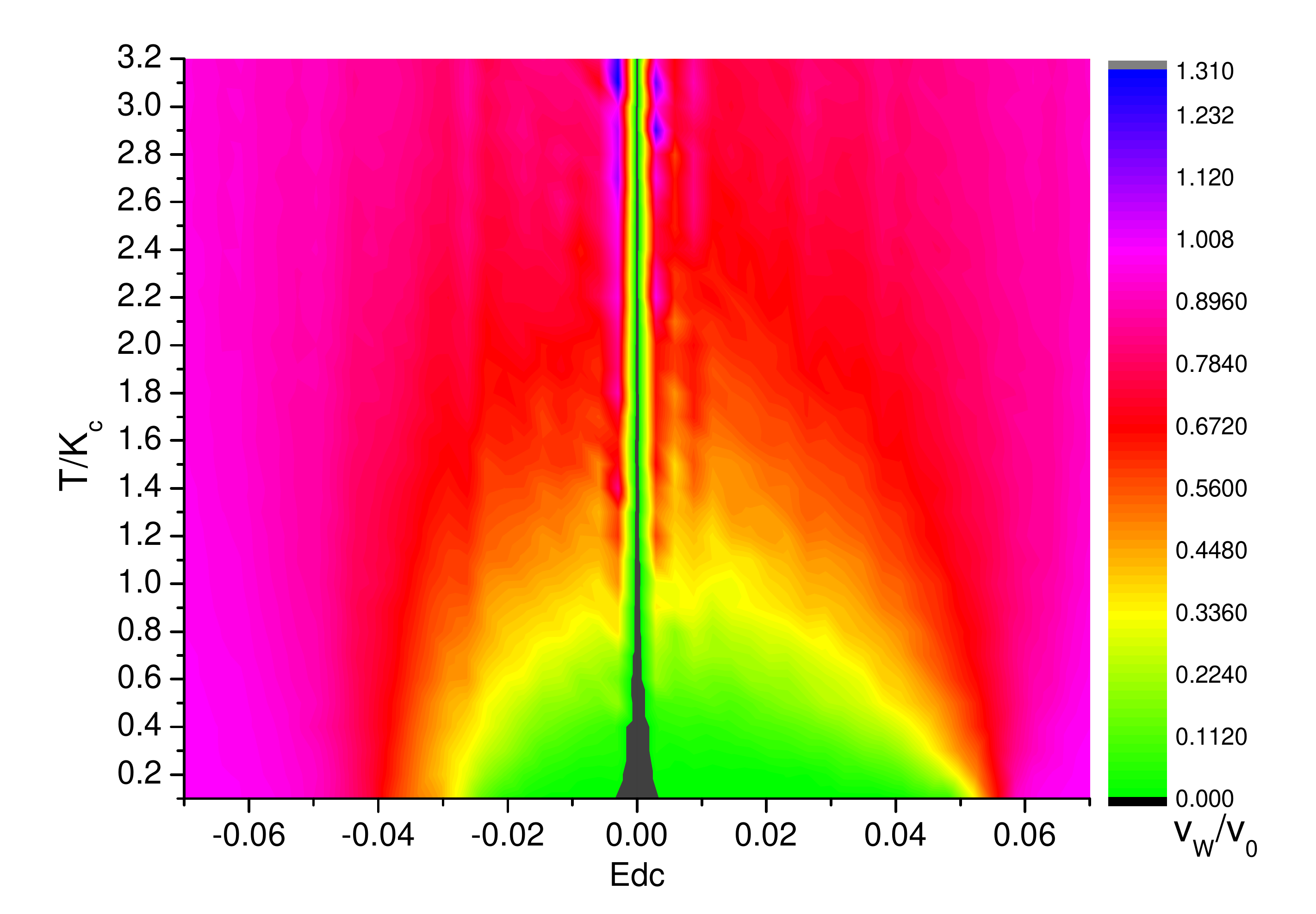}
\end{center}
\vskip -0.3cm
\caption{\label{fig9}
Dependence of $v_W/v_0$, shown by color, on $T/K_c$ and $E_{dc}$ 
for 2D case with 5 stripes and $N/L=34/21$;
$K=0.05$, $K_c=0.0462$; $\eta=0.1$, 
$v_0=E_{dc}/\eta$.
}
\end{figure}

The global dependence of  2D Wigner crystal velocity $\mid v_W \mid$
on temperature $T$ and $E_{dc}$ is shown in Fig.~\ref{fig9}
for density $\nu=N/L=34/21$. We find that the diode transport
is well visible for temperatures being smaller than the potential height
$T/K_c \approx T/K < 1$ where $v_W$ is significantly smaller than
$v_0 = E_{dc}/\eta$. At large $T/K_c$ 
values $v_W$ becomes close to $v_0$ since in this regime
the potential influence becomes small.
Of course, at very small $E_{dc}$ values 
the velocity $v_W$ becomes very small and it becomes difficult
to determine exactly very small $v_W$ values in numerical simulations
on a finite time scale (here $v_W$ appears due to 
exponentially rare thermal fluctuations). 
This is at the origin of peak-like structure 
near $E_{dc} \approx 0$ in Fig.~\ref{fig9}.

Overall, the results for diode transport in 2D
show that its properties are similar to those of 1D case.

We note that using GPUs in the numerical experiments
allows to make simulations with a significantly
larger number of particles going up to
$N_{tot} \sim 10^4$ without a significant increase
of computational times. However, this work was focused
on analysis of specific physical effects
related to diode transport for which 
it was sufficient to stay within a maximal
$N_{tot} =445$ (see below). Recently,
the numerical simulations of 2D Wigner crystal
in a spatially modulated system
with up to 1600 charges has been reported in \cite{dykman}.
However, the possible links with the Aubry transition
have been not discussed in this work.
We think that without such links it is
rather difficult to understand the physics
of various dislocation phases appearing in 2D.

Another thing worth mentioning is the possibility of
realization of thermal diode discussed in \cite{baowenli}.
On the first glance it seems natural
that in presence of charge diode transport one
can expect the thermal diode flow to appear
in our model. However, we were not
able to find thermal diode regime in our model.
We explain this by the fact that the charge diode
transport appears at finite $E_{dc}$ field values.
Such fields are rather moderate on a local scale
but for a large system size they correspond
to a significant voltage difference $\Delta V_{dc}$ 
applied to the sample. In contrast, the temperature
difference applied to the sample is
always smaller than the sample temperature
and due to that to long sample the temperature gradient 
becomes very small so that, due to the detailed balance
\cite{landau10}, the left and right heat flows
remains equal in this linear limit 
of small temperature gradient.

In the next Section we characterize the structure of 
moving Wigner crystal in 2D.

\section{Formfactor of moving Wigner crystal}
\label{sec6}

To characterize the structure of Wigner crystal
in moving and static regimes we us the formfactor
defined as

\begin{eqnarray}
 F(k) =\langle \mid {\it Re} \sum_{i \neq j}^{N_{tot}} \exp(i k (x_i(t)-x_j(t))) \mid^2 \rangle/N_{tot} \; ,
\label{eq:formfactor}
\end{eqnarray} 
where the average is done over all particles and 10 different moments of time 
homogeneously spaced on the whole computational interval of time. 
A similar approach had been used in \cite{fki2007}.
It showed that the formfactor captures the Aubry transition
from sliding phase with $F(k)$ peaks at incommensurate values
$k \approx \nu j$ with integer $j$ while in the pinned phase
the peaks are more pronounced at 
$k \approx j$ corresponding to the lattice period
(see Fig.~5 in \cite{fki2007}).
Of course, in the pinned phase the density is still 
incommensurate (e.g. $\nu =1.618...$) but
the  charges are clustered in groups where their 
positions are more closely located
to potential minima with some displacements (dislocations)
between clusters due to fractal devil's staircase structure
of the whole chain in the pinned phase.
These clusters give peaks of $F(k)$ at $k \approx j$.

\begin{figure}[t]
\begin{center}
\includegraphics[width=0.49\textwidth]{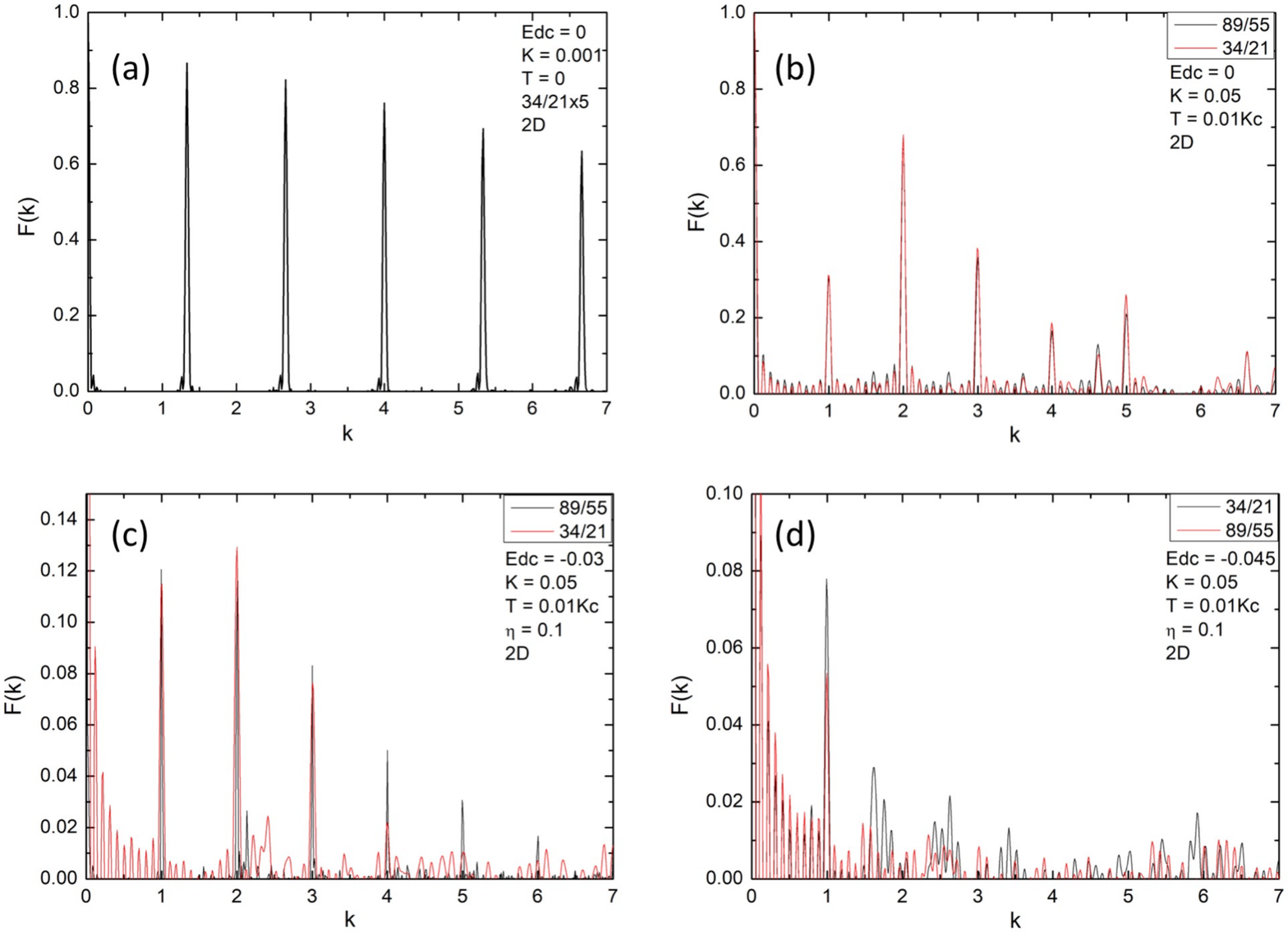}
\end{center}
\vskip -0.3cm
\caption{\label{fig10}
Formfactor $F(k)$ in 2D case with 5 stripes for
(a) $N/L=34/21$; $K=0.001$, $T=0$
($\eta$ is not important since $E_{dc}=0$);
(b) $N/L=34/21; 89/55$, $K=0.05$, $T=0.01Kc$ 
($\eta$ is not important since $E_{dc}=0$);
(c) same as (b) but at $E_{dc}=-0.03$, $\eta=0.1$;
(d) same as (b) but at $E_{dc}=-0.045$, $\eta=0.1$;
}
\end{figure}

For the static Wigner crystal at $E_{dc}=0$ we also find a
similar $F(k)$ structure shown in Fig.~\ref{fig10}(a)
and Fig.~\ref{fig10}(b) for sliding and pinned
phases respectively. However, in the sliding phase at $K=0.001 \ll K_c$
we have $F(k)$ peaks located at $k \approx \nu_{eff} j$ with 
$\nu_{eff} \approx \sqrt{\nu} \approx 1.272$. Indeed, in 2D 
the average distance between particles becomes $1/\sqrt{\nu}$
instead of $1/\nu$ as it was in 1D case. Due to that 
we find peaks at $k \approx  \sqrt{\nu} j$ in Fig.~\ref{fig10}(a).
In contrast, in the pinned phase the periodic potential imposes peaks
at $k \approx j$ in Fig.~\ref{fig10}(b).

The formfactor of the moving Wigner crystal in the diode regime
at $E_{dc}=-0.03, -0.045$ is shown respectively in Fig.~\ref{fig10}(c),(d)
for the pinned phase. The main peaks are still located at integer
$k \approx j$ even though they are somewhat broaden due to fluctuations
of changes during their propagation along the lattice. 
These fluctuations mainly affect large peaks at large $j =3,4 ...$
but the peaks at $j=1,2$ remain. 
It is important to note that with the increase of the ring size
going from $N/L=34/21$ to $N/L=89/55$ we recover the same formfactor structure.
This shows that the chosen system size corresponds to the 
thermodynamic limit of infinite system.

\begin{figure}[t]
\begin{center}
\includegraphics[width=0.49\textwidth]{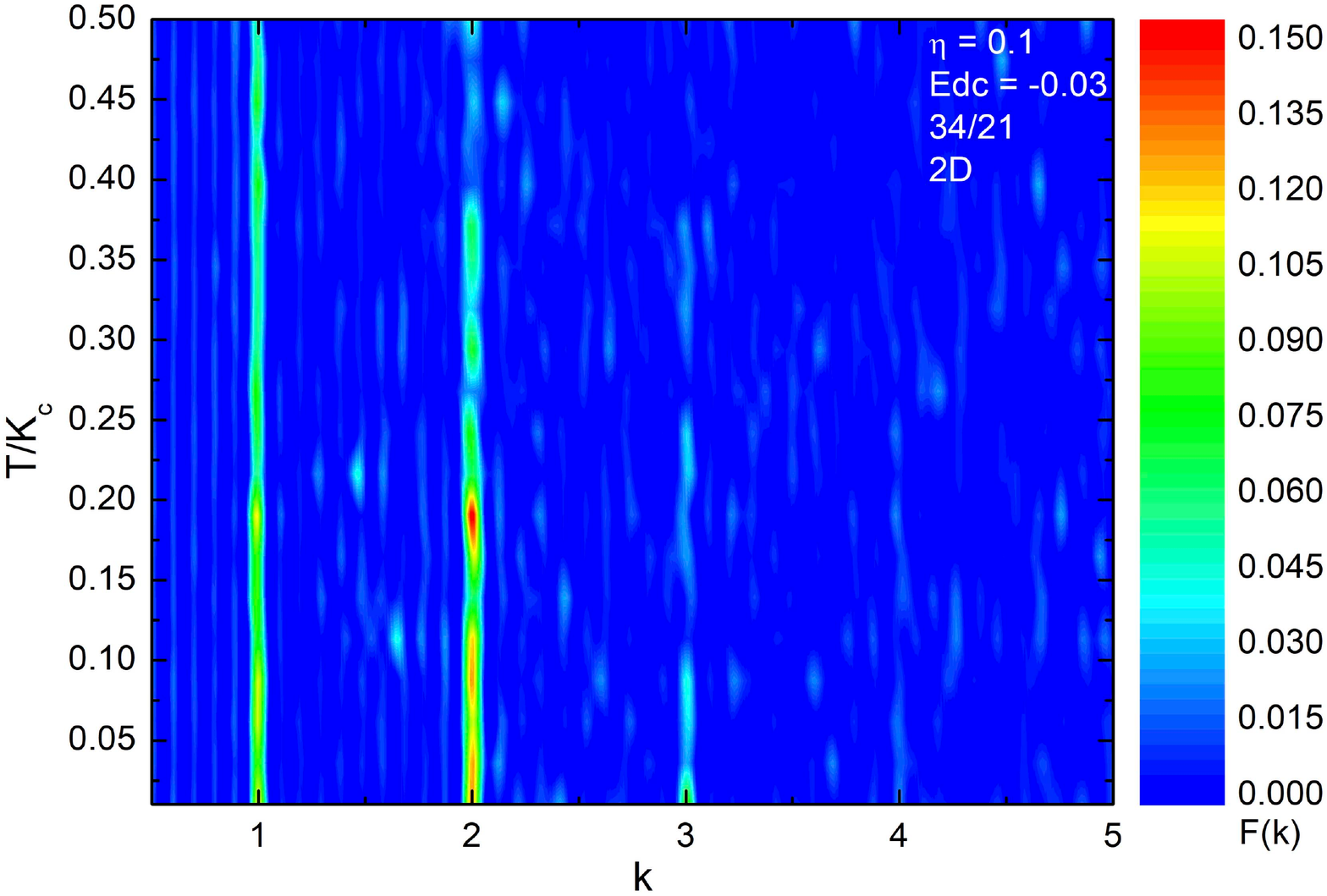}
\end{center}
\vskip -0.3cm
\caption{\label{fig11}
Dependence of formfactor $F(k)$, shown by color, on
$T/K_c$ and $k$; $N/L=34/21$, $\eta=0.1$,
$E_{dc}=-0.03$ in 2D with 5 stripes.
}
\end{figure}

The dependence of $F(k)$ on temperature $T$
for moving crystal at $E_{dc}=-0.03$ is shown in Fig.~\ref{fig11}.
With the increase of $T$ the fluctuations become stronger
and the peaks are suppressed at large $T/K_c \approx T/K$ values.
However, the peaks at $k \approx j =1,2$ remain rather robust
even at large $T/K$.

\begin{figure}[t]
\begin{center}
\includegraphics[width=0.49\textwidth]{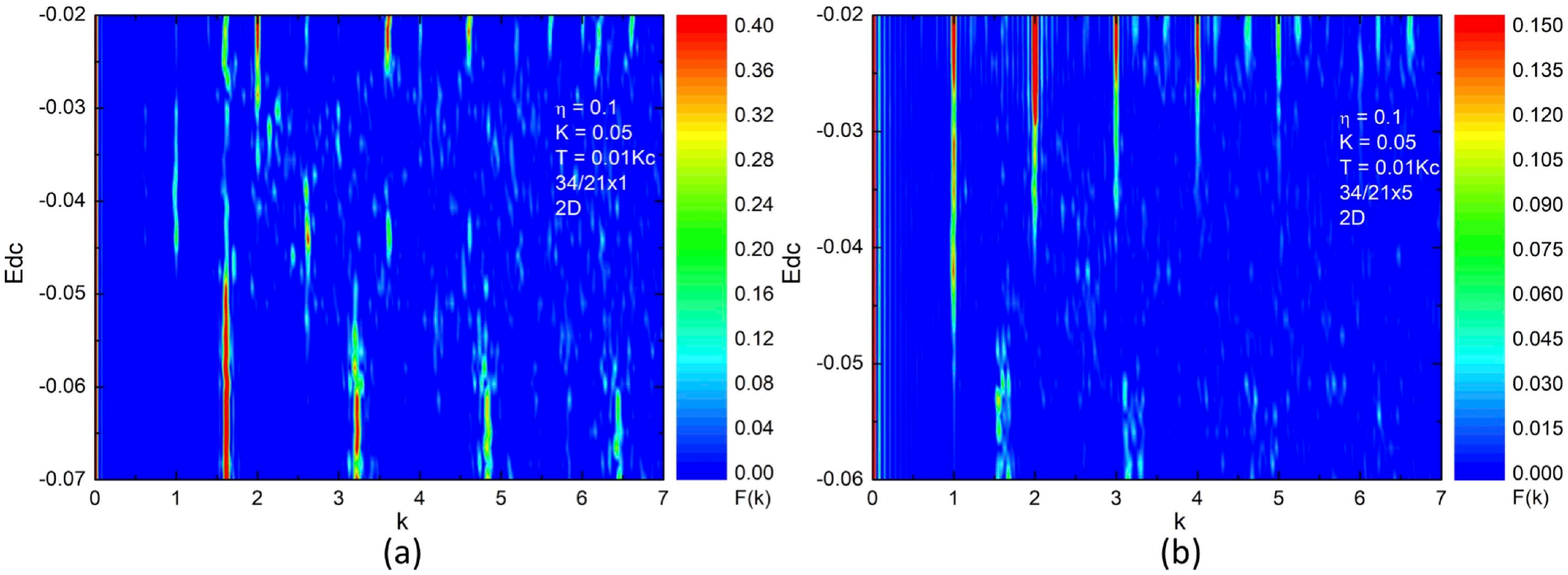}
\end{center}
\vskip -0.3cm
\caption{\label{fig12}
Dependence of formfactor $F(k)$ of moving Wigner crystal
on $E_{dc}$ and $k$ for 
$K=0.05$, $T=0.01K_c$, $K_c=0.0462$, $\eta=0.1$ in 2D at
$N/L=34/21$ with 1 stripe (a)
and with 5 stripes (b) 
(color bar marks $F(k)$ values). 
}
\end{figure}

The dependence of formfactor on $E_{dc}$ 
at fixed temperature is shown in Fig.~\ref{fig12}.
At moderate $\mid E_{dc} \mid < 0.05$ 
values in the diode regime we have $F(k)$ peaks
mainly at $k \approx j =1,2,3,4,5,6$
(they are more pronounced for large size
with 5 stripes). But at stronger field $\mid E_{dc} \mid >0.05$
we see a transition to incommensurate structure
with peaks located at $k \approx {\tilde \nu} j$
with $ {\tilde \nu} \approx 1.5$ corresponding to 
an intermediate density between $\nu_{eff} \approx 1.272$ and
$\nu =1.618$. Thus a sufficiently strong {\it dc-}field 
can change the structure of moving pinned Wigner crystal
to sliding incommensurate crystal. Indeed, at
strong $E_{dc}$ the  crystal velocity $v_W$ becomes rather 
large and the periodic potential gives to this 
directed flow only relatively weak perturbation. 

We present the video of Wigner crystal motion
in the diode regime in Supplementary Material \cite{supmatvideo}.

\section{Analysis of approximations}
\label{sec7}

The performed numerical simulations
use certain approximations like e.g. 
the Langevin approach for the thermal bath,
finite interaction range between charges.
Here we briefly justify the validity
of these approximations.

Thus, in Fig.~\ref{figA4} we show the velocity distributions
in 1D and 2D cases. These results show that the numerical
results well reproduce the theoretical thermal Maxwell distribution
centered at the average particle velocity $v_W$.
At significant values of $E_{dc}$ and related $v_W$ 
there is a noticeable deformation of the distribution
which we attribute to a moderate heating of particles
by the static force $E_{dc}$ and effects of their interactions
in presence of a periodic potential.
However, we note that even when there is a visible deformation
of the distribution in $x$-direction in 2D case,
the distribution in $y-$direction is still very 
well described by the Maxwell one. We note that as it was described
Section~\ref{sec2} in 2D case the particles are free
to move from one stripe to another one but such transitions
happened to be rare due to repulsion between charges.

We note that the friction force for electrons on a surface of liquid helium
can have a more complicated form due to various
low energy excitations in this system (see e.g. \cite{konobook}).
However, the Langevin approach reproduces well the thermal distribution
in the system and we consider that this approach is reasonably
justified as the first step to investigation of 
this rather complex and nontrivial system.
However, we suppose that the future studies
will allow to test other forms of the friction force
in the Langevin equation that will capture the specific
features of electron friction on the surface of liquid helium
due to rich properties of low energy
excitations in this system.

Indeed, the results presented in Fig.~\ref{figA5}
show that the temperature 
dependence of Wigner crystal velocity $v_W$
is well described by the  Arrhenius thermal activation 
equation that confirms that the Langevin approach
provides us a reasonable description of the thermal
environment.

The obtained results for $v_W$ are not sensitive to the interaction
radius between charges $R_C$ as it is illustrated in fig.~\ref{figA6}
where it is changed by a factor 50 with practically the same
profile of time dependence $v_W(t)$, up to statistical fluctuations.
This is in agreement with the results of Fig.~\ref{fig2}
which shows that the chain of charges, where all interactions are taken into account,
is rather well described by the symplectic map with only nearby interactions between charges.
We also note that the  comparison between 1D chain dynamics
with only nearest neighbors interactions
and the chain with interactions of all charges had been performed in
\cite{fki2007,ztzs} showing that the short range approximation
for interactions provides rather good approximation.
In addition, we also show in Fig.~\ref{figA7}
that the dependencies $v_W(E_{dc})$ for 1D case
(with only nearby interactions) and 2D case (with 1 stripe/line)
remain very close to each other. Thus 1D case captures the main physical 
features of the system being well useful for the description of 2D system.

Thus, we consider that the results discussed above justify the 
approximations used in our numerical simulations.

\section{Discussion}
\label{sec8}

In this work we demonstrated 
that the Wigner crystal diode transport 
appears naturally for charge motion
in asymmetric 1D and 2D potential.
In presence of charge interactions
a {\it dc-}field   move crystal easily
in one direction while
no current appears in opposite direction.
Our results show close similarities
of diode transport in 1D and 2D.
The diode transport appears in the Aubry pinned phase.

We think that the asymmetry is rather natural 
for various materials since
already three different atoms in a
periodic cell create generally 
an asymmetric potential.
An incommensurate charge density 
in such materials can be induces 
by effective charge doping
from other planes (e.g. like
in  high-temperature cuprates superconductors)
or impurities.  

With a recent progress in experimental studies
of Aubry transition with cold ions 
\cite{vuletic2016natmat,ions2017natcom}
and colloidal monolayers \cite{bechingerprx}
we  hope to obtain a deeper understanding of
mechanisms of nanofriction on atomic scale \cite{tosatti2}.
These experiments can be also performed 
with asymmetric potentials 
providing first experimental realizations 
of Wigner crystal diode. Indeed, two-harmonic
optical lattices had been already realized experimentally
(see e.g.~\cite{inguscio,bloch}) that opens
possibilities to study the diode regime with cold ions.
The dependence of the Aubry transition
on density, obtained in \cite{fki2007}
(see (\ref{eq:kc}) and \cite{ztions}), shows that 
the Aurby phase can be reached with a moderate
amplitude of lattice potential created by
laser fields. Indeed, for $\nu \approx 0.38$
we obtain from (\ref{eq:kc}) $K_{cv} \approx 0.00044$ 
with the required potential amplitude of Aubry transition being
$V_A=K_{c\nu} e^2/(\ell/2\pi) \approx 0.04 K^{\circ} (Kelvin) $  for the lattice period
$\ell \sim 1 \mu m$ ($V_A \approx 3 K^{\circ}$ for $\nu =1.618$). 

In addition to cold ion experiments we think that 
there are promising possibilities to study Wigner crystal diode
with electrons on liquid helium moving in a
quasi-1D channel \cite{kono1d}. The first experiments in this direction
have been  reported recently in \cite{konstantinov}. 
According to the above estimate, for electron density
$\nu \approx 1.618$ and $\ell \approx 1 \mu m$
the Aubry transition takes place at the potential amplitude
$V_A \approx 3 K^{\circ}$ that is well below the electron
temperature of about $0.1 K^{\circ}$ well available
for such experiments (see e.g. \cite{kono1d,konstantinov}).

There are also possibilities of experimental
realization of Wigner crystal diode
with colloidal monolayers extending the experiments \cite{bechingerprx}
to asymmetric potentials. 

We expect that the experimental investigations of
electron and ion transport in a periodic potential
at low temperatures will allow to understand the
nontrivial mechanisms of friction and thermoelectricity
at nanoscale and then on atomic scale
with new applications for material science.

\vspace{1cm}
\textit{Acknowledgments.--} We thank N.~Beysengulov,
A.D.~Chepelianskii, J.~Lages, D.A.~Tayurskii and O.V.~Zhirov 
for useful remarks and discussions.
This work was supported in 
part by the Programme Investissements
d'Avenir ANR-11-IDEX-0002-02, 
reference ANR-10-LABX-0037-NEXT (project THETRACOM).
This work was granted access to the HPC GPU resources of 
CALMIP (Toulouse) under the allocation 2018-P0110. 
The development of the VexCL library was partially 
funded by the state assignment
to the Joint supercomputer center of 
the Russian Academy of sciences for scientific
research. The work of M.Y. Zakharov was partially funded 
by the subsidy allocated to Kazan Federal
University for the state assignment in 
the sphere of scientific activities (project
$N^{\circ}$ 3.9779.2017/8.9).

\appendix 
\section{}
\label{appenda}
\setcounter{figure}{0} \renewcommand{\thefigure}{A.\arabic{figure}} 

\begin{figure*}[!htb]
\begin{center}
\includegraphics[width=0.9\textwidth]{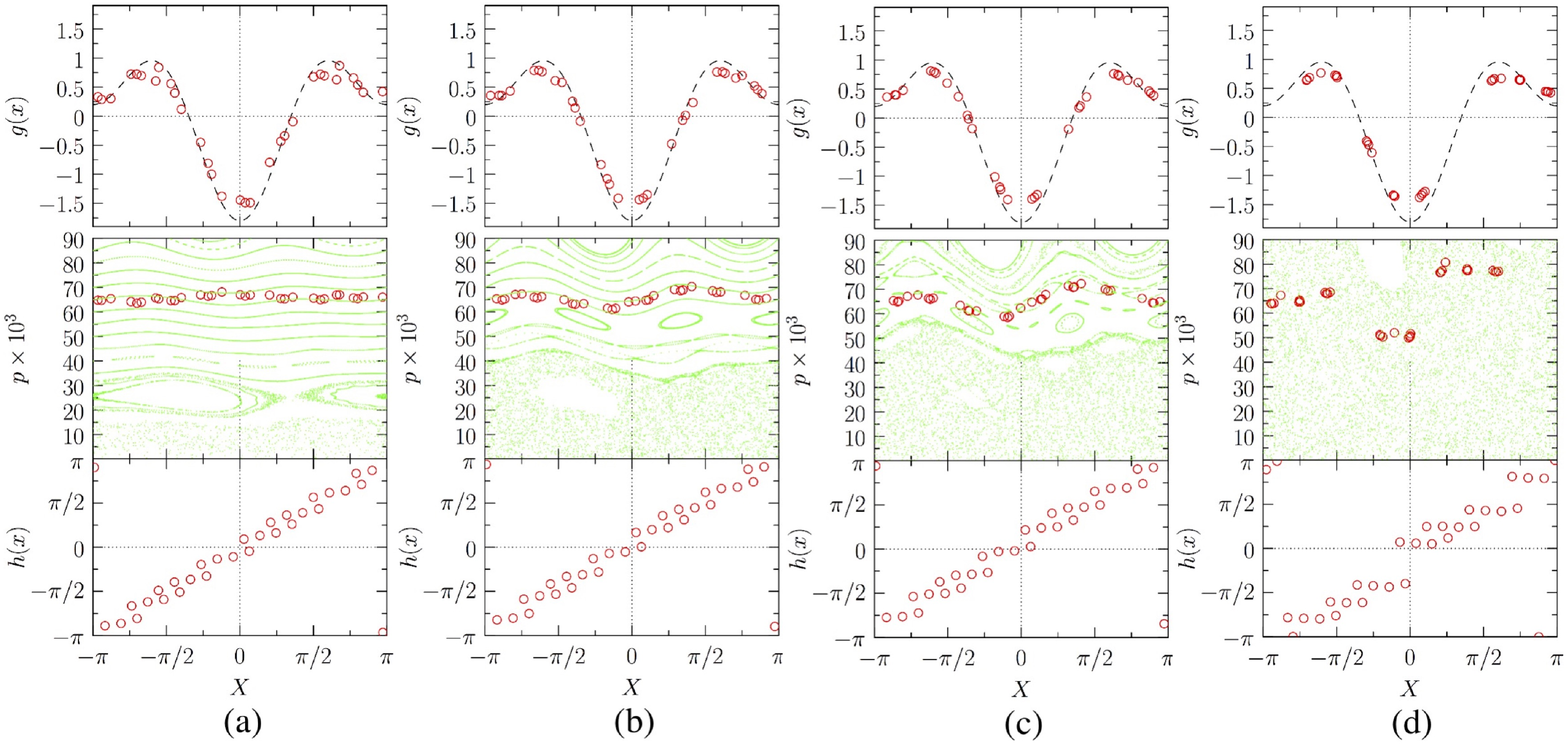}
\end{center}
\caption{\label{figA1}
Same as in Fig.~\ref{fig2} but for
$\nu \approx 1.618$ with $N=89$ charges for $L=51$ potential periods
and $K=0.002$ (a), $K=0.005$ (b), $K=0.008$ (c), $K=0.02$ (d). 
}
\end{figure*}

\begin{figure}[!htb]
\begin{center}
\includegraphics[width=0.48\textwidth]{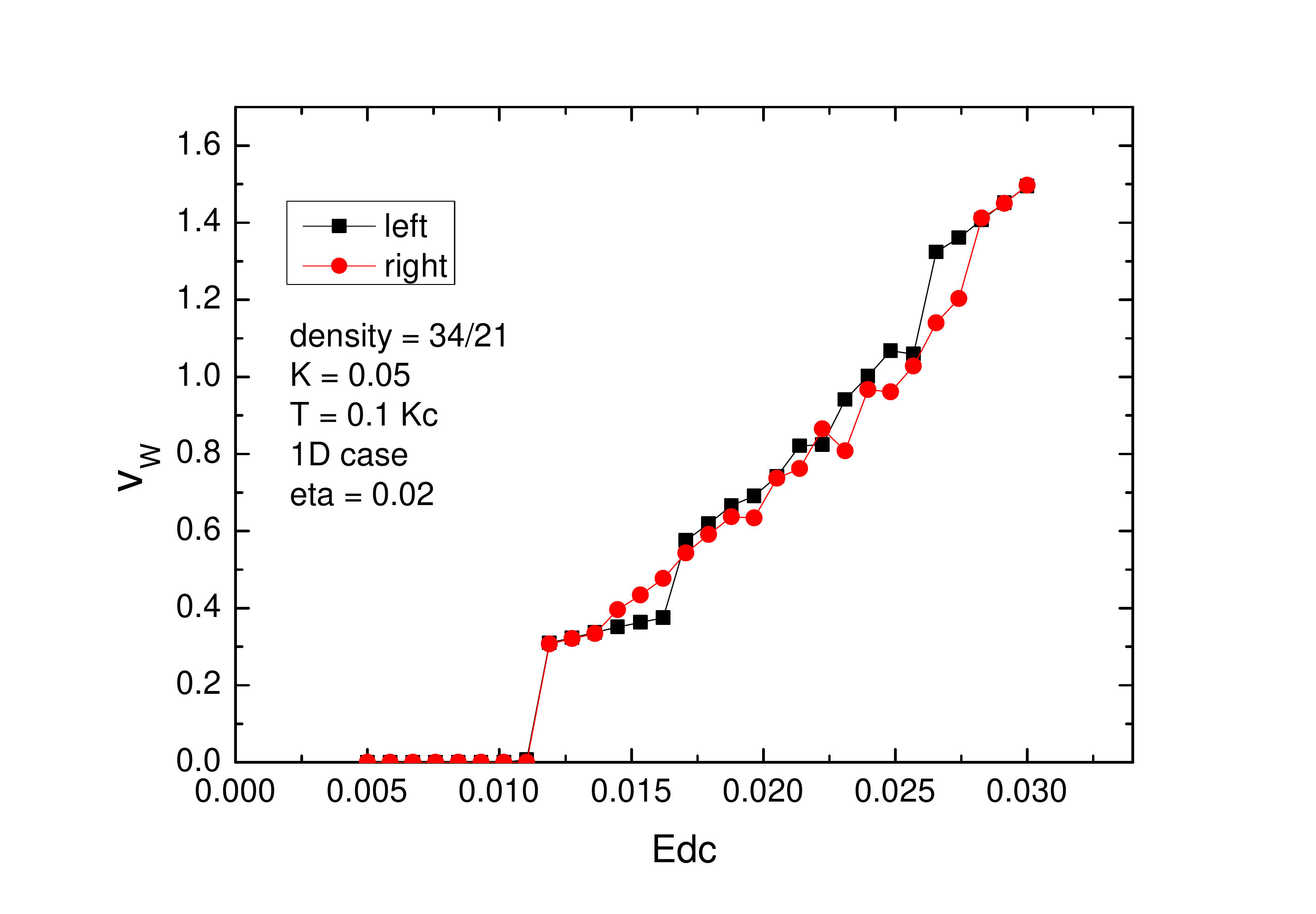}
\end{center}
\caption{\label{figA2}
Same as in Fig.\ref{fig7}(b) 
but at $\eta=0.02$ (instead of $\eta=0.1$).
}
\end{figure}

\begin{figure}[!htb]
\begin{center}
\includegraphics[width=0.48\textwidth]{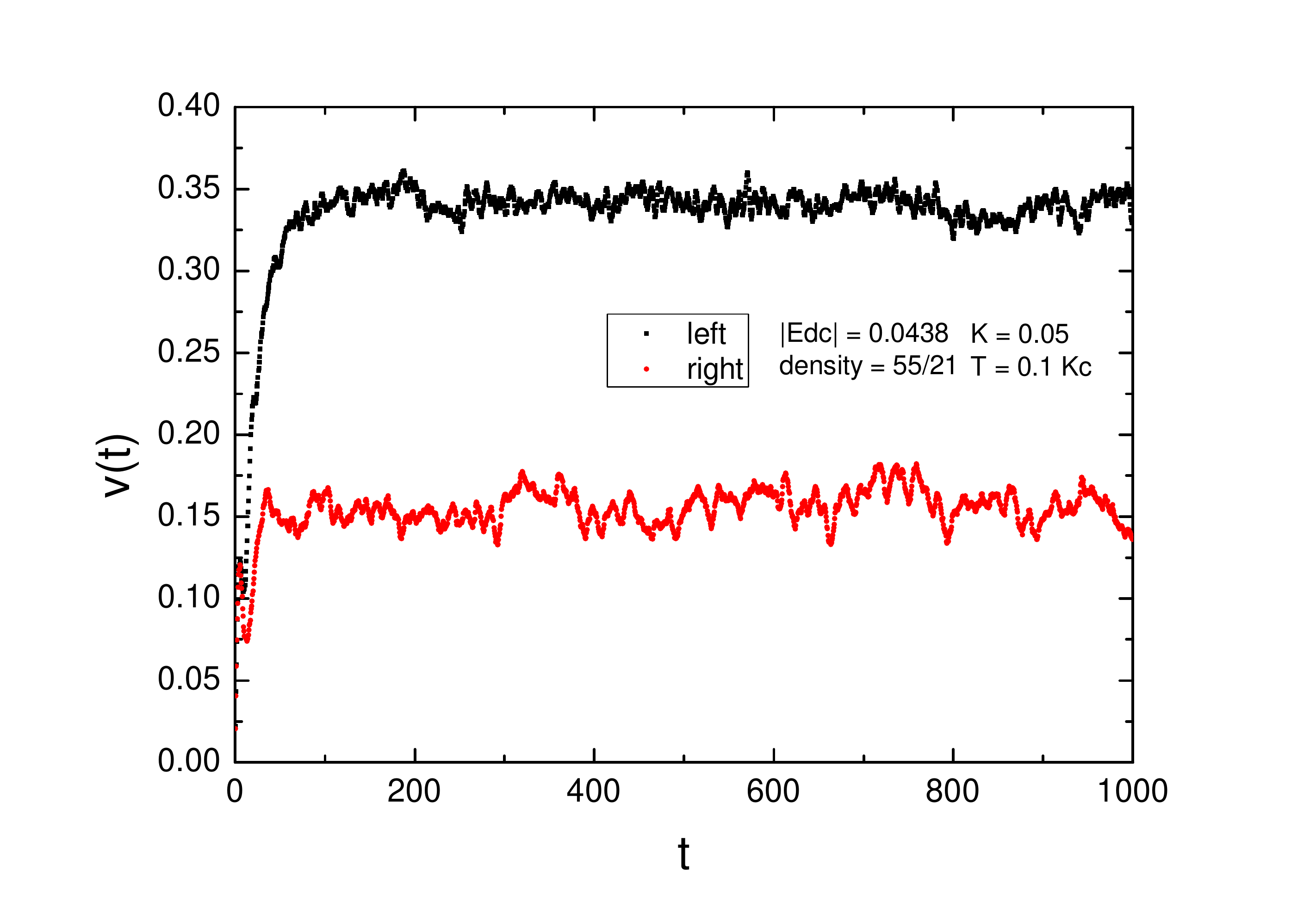}
\end{center}
\caption{\label{figA3}
Time dependence of  velocity in $x$ in 2D 
with $N_s=5$ stripes; here $\nu =N/L= 55/21$; 
$K=0.05$, $T=0.1K_c=0.00462$, $\eta=0.1$, $\mid E_{dc} \mid =0.0438$.
The time averaged Wigner crystal velocities are
$v_W=v_{left}=-0.34$ and $v_W=v_{right}=0.16$
}
\end{figure}

\begin{figure}[!htb]
\begin{center}
\includegraphics[width=0.48\textwidth]{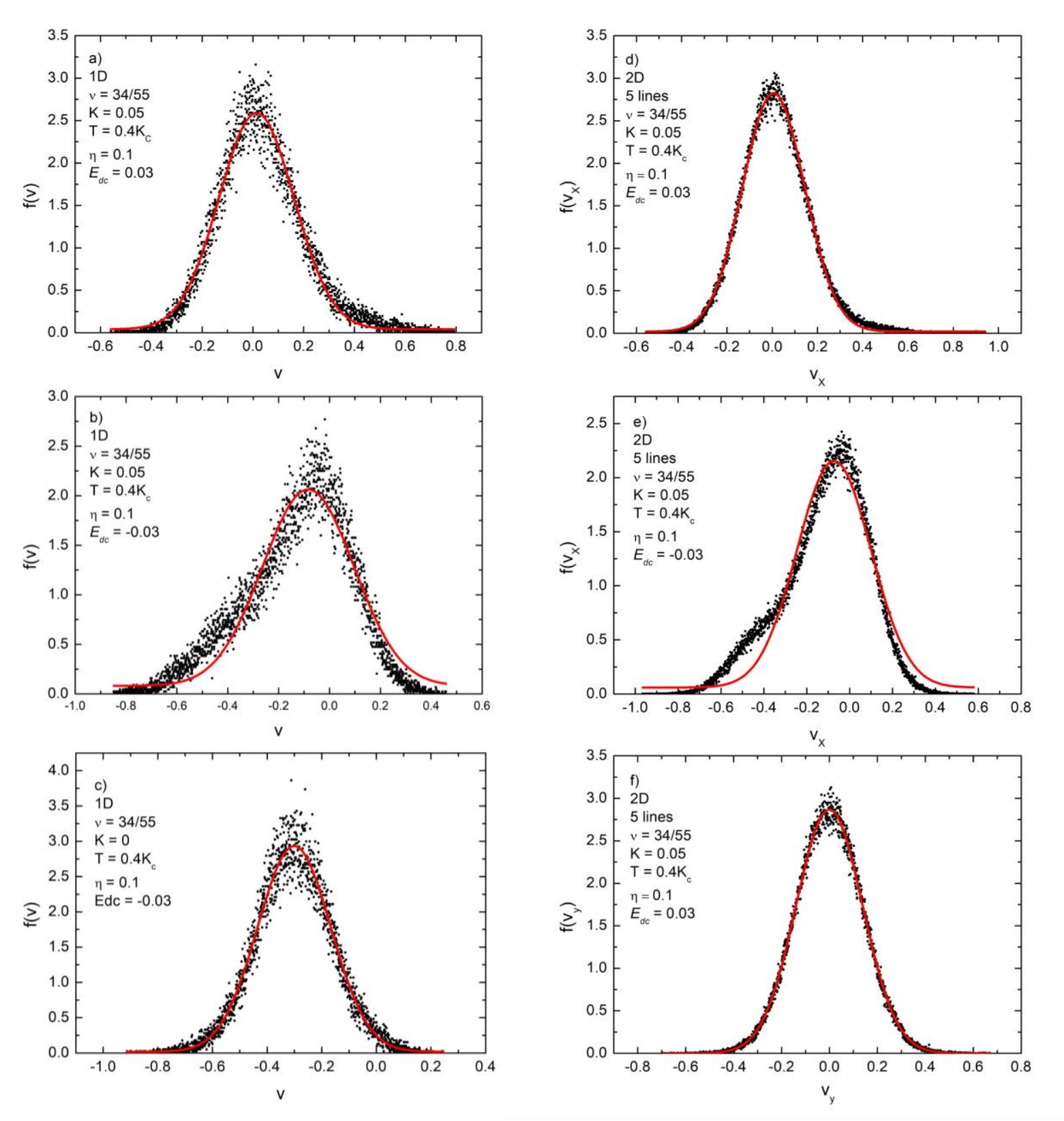}
\end{center}
\caption{\label{figA4}
Verification of the thermalization of charges
in 1D and 2D (5 stripes/lines), black points show particle velocities 
in the time range  $200 \leq t \leq 300$, the red curve shows
the theoretical thermal Maxwell distribution centered at the average
velocity of particles $v_{W}$ in $x-$direction. System parameters are 
given in figure panels.
}
\end{figure}

\begin{figure}[!htb]
\begin{center}
\includegraphics[width=0.48\textwidth]{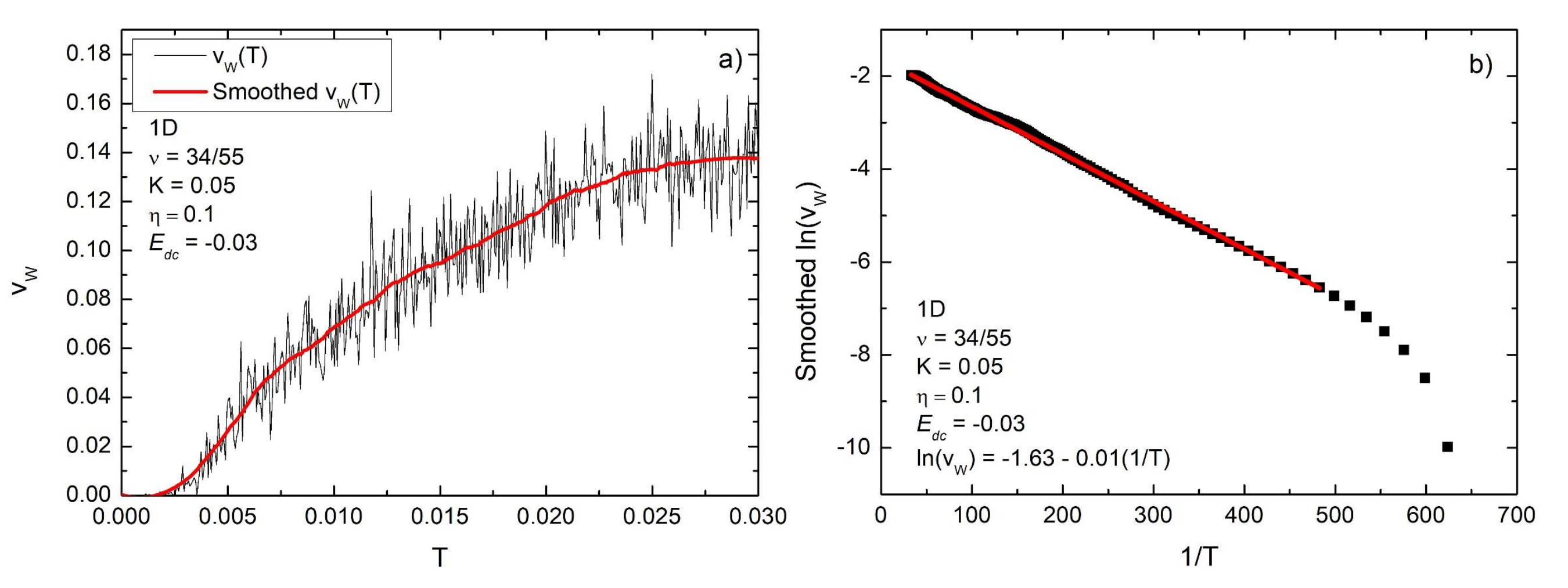}
\end{center}
\caption{\label{figA5}
Dependence of the Wigner crystal velocity
$v_W$ on temperature $T$  in 1D case;
panel (a): numerical results for $v_W$
are shown by small points connected by the black curve,
red curve shows the smoothed dependence obtained 
by the Savitzky-Golay filter with polynomial order 2 
(points of window 50 in ORIGIN package); panel (b): black squares
show data for the smoothed red curve of panel (a),
the red line shows the Arrhenius thermal activation dependence
with the fit parameters $\ln v_W = -0.163 - A_r/T$ 
at the activation energy $A_r=0.01$.  The system parameters are 
given in the panels. 
}
\end{figure}

\begin{figure}[!htb]
\begin{center}
\includegraphics[width=0.48\textwidth]{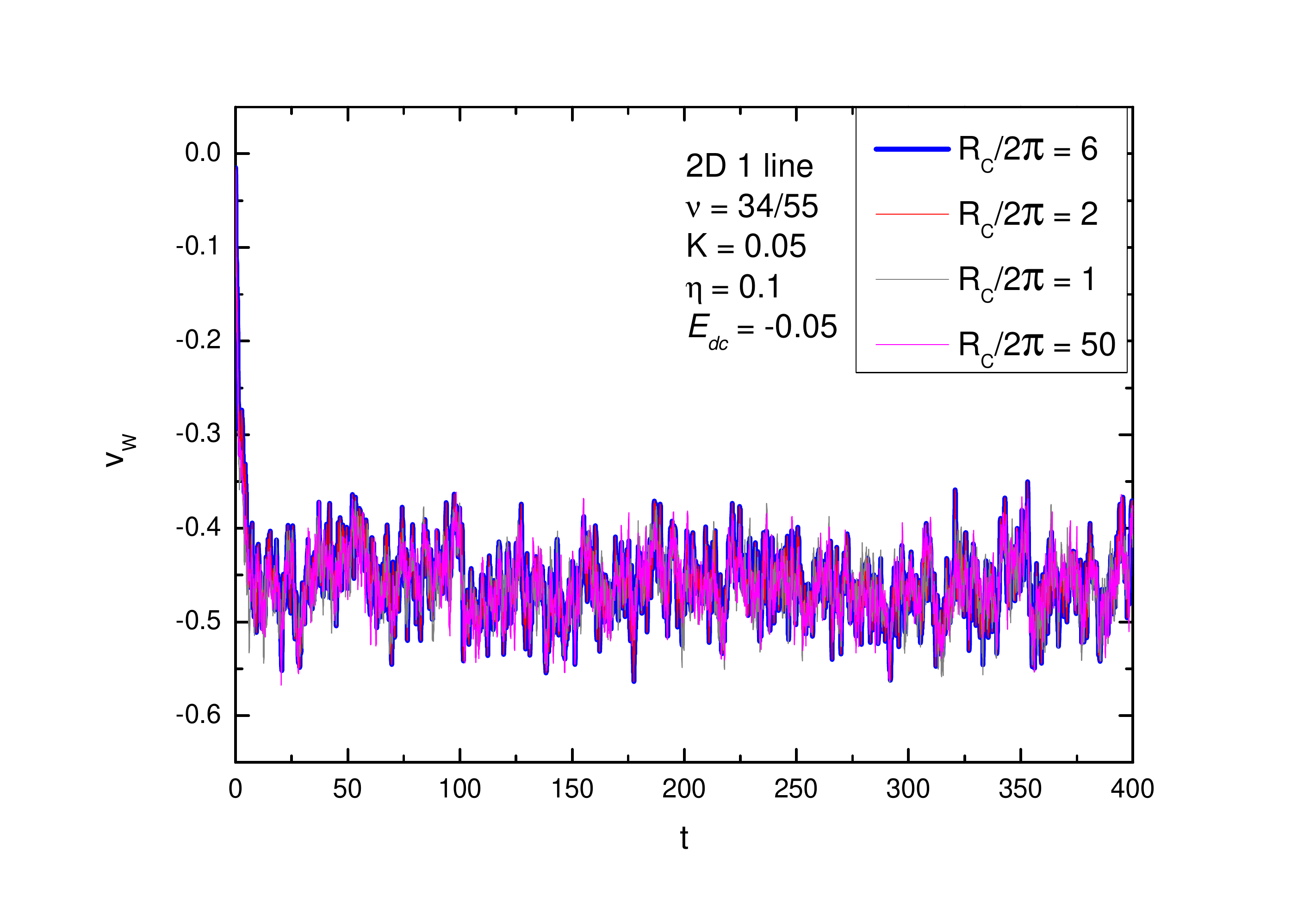}
\end{center}
\caption{\label{figA6}
Dependence of $v_W$ on time $t$
in 2D case (1 stripe/line, velocity is in $x-$direction )
for different values of interaction radius
$R_C/2\pi = 1, 2, 6$ (our main case), $50$;
system parameters are given in the panel.
}
\end{figure}

\begin{figure}[!htb]
\begin{center}
\includegraphics[width=0.48\textwidth]{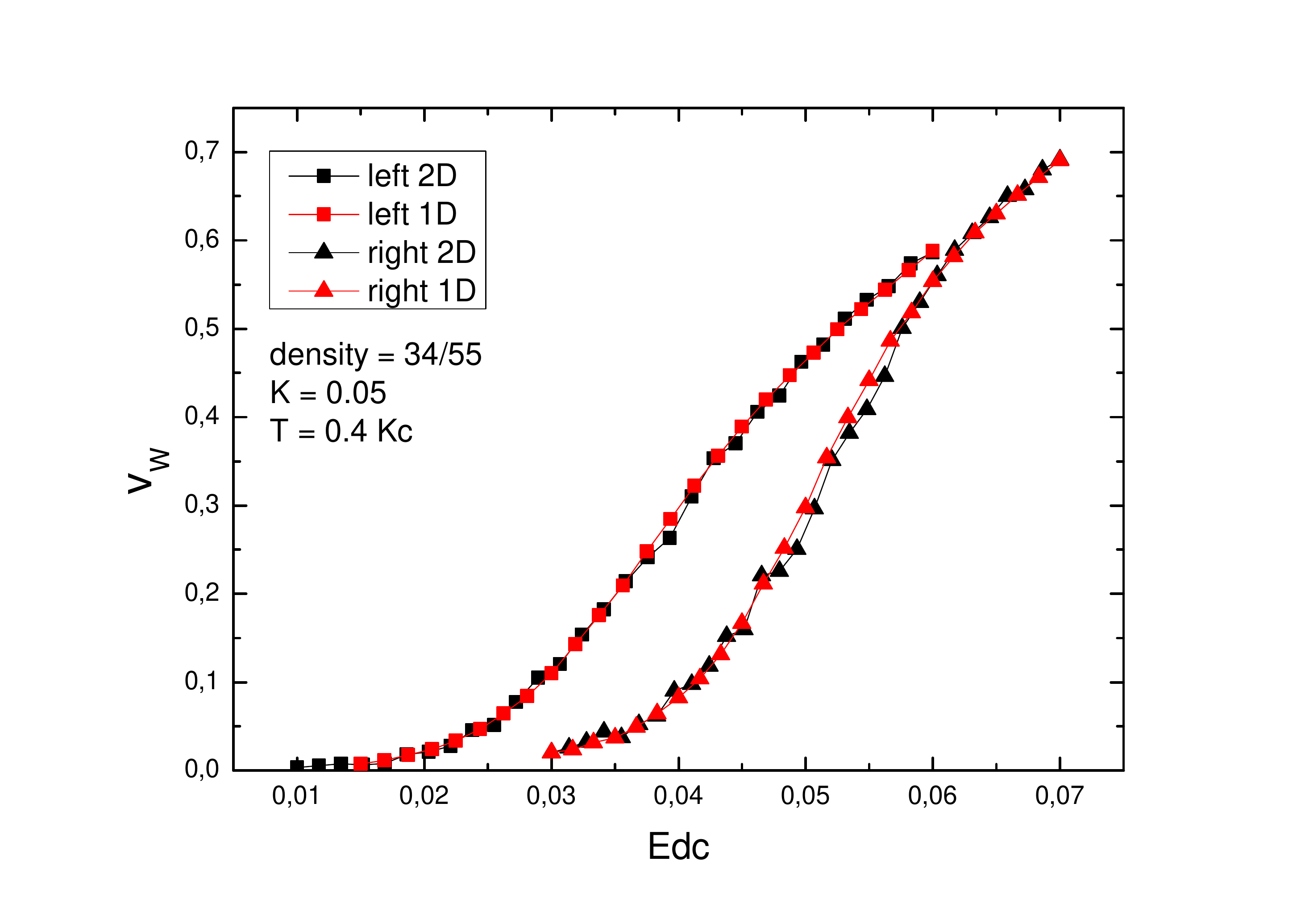}
\end{center}
\caption{\label{figA7}
Dependence of $v_W$ on $E_{dc}$ for
1D case and 2D case 
(1 stripe/line; $R_C/2\pi = 6$, velocity is in $x$-direction);
system parameters are given in the figure panel.
}
\end{figure}

Here we present supplementary Appendix 
figures Fig.~\ref{figA1}, Fig.~\ref{figA2}, Fig.~\ref{figA3},
Fig.~\ref{figA4}, Fig.~\ref{figA5}, Fig.~\ref{figA6}, Fig.~\ref{figA7}
complementing
the main text of the paper.


\end{document}